\documentclass[11pt]{article}
\usepackage{epsfig}
\usepackage{graphicx}
\usepackage{amssymb}
\usepackage{here}
\usepackage{cite}
\newcommand{\AmS}{{\protect\the\textfont2
  A\kern-.1667em\lower.5ex\hbox{M}\kern-.125emS}}
\newcommand{\ba}{\begin{array}}
\newcommand{\ea}{\end{array}}

\def\beq{\begin{equation}}   
\def\eeq{\end{equation}}

\def\bea{\begin{eqnarray}}
\def\eea{\end{eqnarray}}

\def\beq{\begin{equation}}   
\def\eeq{\end{equation}}

\def\bea{\begin{eqnarray}}
\def\eea{\end{eqnarray}}

\begin{document}
\begin{titlepage}

\begin{flushright}
Report: IFIC/08-61, FTUV-08-1212\\
\end{flushright}

\begin{center}
\vspace{2.7cm}
{\Large{\bf Prospects of searching for (un)particles from Hidden Sectors}}
\end{center}
\begin{center}
{\Large {\bf using rapidity correlations 
in multiparticle production at the LHC}}
\end{center}

\vspace{1cm}

\begin{center}

{\bf Miguel-Angel
Sanchis-Lozano
\footnote{Email:Miguel.Angel.Sanchis@ific.uv.es}
\vspace{1.5cm}\\
\it 

\it Instituto de F\'{\i}sica
Corpuscular (IFIC) and Departamento de F\'{\i}sica Te\'orica \\
\it Centro Mixto Universitat de Val\`encia-CSIC, 
Dr. Moliner 50, E-46100 Burjassot, Valencia (Spain)}

\end{center}

\vspace{0.5cm}

\begin{abstract}

\end{abstract}
Most signatures of new physics have been studied on the
transverse plane with respect to the beam direction
at the LHC where background is much reduced. 
In this paper we propose 
the analysis of inclusive longitudinal (pseudo)rapidity correlations 
among final-state (charged) particles in order to search 
for (un)particles belonging to a Hidden Sector
beyond the Standard Model, using a selected sample of p-p minimum bias events 
(applying appropriate off-line cuts on events based on, e.g.,
minijets, high-multiplicity, event shape variables, high-$p_{\bot}$ leptons 
and photons, etc) collected at the early running of the LHC. 
To this aim, 
we examine inclusive and semi-inclusive two-particle
correlation functions, forward-backward correlations, and
factorial moments of the multiplicity distribution, 
without resorting to any particular model but
under very general (though simplifying) assumptions. Finally, motivated by
some analysis techniques employed in the search for Quark-Gluon-Plasma
in heavy-ion collisions, we
investigate the impact of such intermediate (un)particle stuff
on the (multi)fractality of parton cascades in p-p collisions, 
by means of a L\'evy stable law description and a  
Ginzburg-Landau model of phase transitions. Results from our
preliminary study seem encouraging for possible dedicated 
analyses at LHC and Tevatron experiments.
\\
\vspace{1cm} 

\begin{center}


\end{center}

\end{titlepage}


\tableofcontents

\newpage

\section{Introduction}

Most signatures and signals 
of new phenomena are expected to be found 
in hard events at the LHC
on the {\em transverse} plane with respect to
the beam direction where new physics (NP)
would show up while background is much reduced. 
Typical examples for $\lq\lq$direct discovery'' 
are high-$p_{\bot}$ jets, large missing $E_{\bot}$,
large invariant mass of dileptons, diphotons or
dijets, high-energy monochromatic photons, etc (see 
e.g. \cite{Mangano:2008ag} and references
therein). However, a proposal 
has been put forward recently  
\cite{Harnik:2008ax} to search for new physics
in {\em anomalous underlying events}, looking for soft diffuse signals 
in the tracking system and calorimetry of the LHC experiments.
Earlier, possible signals of NP from global event
properties (notably manifesting in the tail of the charged 
particle multiplicity distribution)
were studied in \cite{Giovannini:2003ft}.

In this paper we propose to the CERN LHC collaborations
\cite{lhc,alice} 
to consider further signatures of NP in multiparticle production, 
by carefully scrutinizing
inclusive (longitudinal) rapidity correlations between 
(charged) particles emitted in inelastic proton-proton (p-p)
interactions. We are particularly 
interested in (un)particles
from a Hidden Sector (HS) beyond the Standard Model (SM) 
(sometimes more generally -and thought-provokingly- 
described as $\lq\lq$hidden worlds'' \cite{Wells:2008xg}) contributing 
through a new state of matter to the hadronization chain
yielding final-state SM particles.  
Our approach does not rely on any particular model, besides
some general (but simplifying) assumptions on hadroproduction,
and should be considered as a starting point for a much more complete 
study.

The study of inclusive densities of secondaries
and correlations between them has indeed been providing for decades
(in cosmic rays physics, fixed target and collider experiments)
extremely useful information about the strong interaction 
dynamics in high-energy interactions. 
Notably, high-order correlations among
emitted particles have been found in all types of hadron
collisions, manifesting through multiplicity fluctuations
when phase-space is split in small cells. The normalized moments 
of the multiplicity distributions then exhibit 
a power-law dependence ($\lq\lq$intermittency''), analogous to that
observed in other (multi)fractal phenomena in Nature, like
galaxy clustering \cite{Jones:2004dv}. Besides, particle
correlations provide essential information on the
properties of jets, with the advantage that the 
systematic error due to missing particles in the jet 
is not so relevant as in other observables as stressed in
\cite{Tannenbaum:2005by}. Actually, 
particle production inside QCD jets follows a self-similar
multifractal pattern, as could be expected from a multiplicative
branching process. 

In this work we argue that an extra non-standard state of matter 
at the onset of the parton cascade might 
significantly change its multifractality character, advocating
the use of estimators to determine the mono- or multi-fractal character
of the hadroproduction process. A suggestive connection 
will be established between our prospects of looking for 
hidden (un)particles in p-p collisions, 
and the formation of quark-gluon 
plasma in heavy-ion collisions, where a L\'evy stable law
description and a Ginzburg-Landau model of phase transitions
have been thoroughly applied to 
study the deconfinement phase transition
\cite{Hwa:1992uq}.

The layout of the paper is as follows. In section 2 we 
present a survey on multibody production  
in inelastic p-p collisions, 
where widely used concepts and tools particularly
important for our proposal are highlighted.
Hidden sectors (Unparticle, Hidden Valley, and other
scenarios like Technicolor, Quirks or even mini Black Holes) 
are very briefly addressed in section 3 regarding
the phenomenology at hadron colliders.
In section 4 we introduce our notation 
and basic formulas for the inclusive
analysis. Cascades
in hadronic collisions are considered in section 5
under the working hypotheses of two- and three-step scenarios,
studying the effect on the correlation functions
and moments of the multiplicity distribution. 
Section 6 is devoted to the study of Forward-Backward correlations.
Intermittency and (multi)fractality 
are examined in section 7 working out an illustrative
$\lq\lq$exercise'' largely motivated by usual analyses of
heavy-ion collisions. 
Finally, a summary and a short proposal for this
kind of analysis at the LHC are given in section 8.

\section{A short survey on multiparticle dynamics}

Experimental results in multibody hadroproduction 
collected along decades have steadily supported 
the tendency of produced particles to merge into 
correlated groups \cite{Dremin:1977wc}. 
This experimental evidence rapidly led to the view 
of a two-step process for high-energy
hadron collisions: More or less well-defined
hadronic objects should emerge during the first stage
of the interaction, subsequently decaying
into final-state particles through a complicated 
and, to a large extent, non-perturbative shower. In
the following \footnote{This section can be 
skipped by the reader already familiar with the topic.}, 
we highlight basic points
of multiparticle dynamics, not in a comprehensive way
at all, but
focusing on those aspects especially related to
our suggestion.
 
\subsection{Early models}
The statistical approach to multiparticle
production was first independently proposed by Fermi and Wilson
in the fifties \cite{Fermi:1950jd,Wilson}. In this
model, the energy of the collision concentrates in a small interaction
volume where various kinds of particles, mainly pions,
are created. A statistical equilibrium was assumed to be
reached in the $\lq\lq$gas'' of such confined particles. The effect of
strong interaction in the original Fermi model was restricted
to establish some sort of statistical equilibrium.
Limited progress of this picture was performed by Landau
\cite{Belenkij:1956cd} in the hydrodynamical model for 
hadroproduction. 

In the Fermi model, the strong interaction
played only a marginal role, defining to some extent the
radius of the initial volume. Strong interactions were taken
more seriously into account by Hagedorn in the
statistical bootstrap model \cite{Hagedorn:1967ua} whereby
resonance-like structures ({\em fireballs}) emerged: 
a fireball consisted of fireballs, in turn
consisting of fireballs, and so on till the hadronization stage. 
The hadronic excitation
spectrum was also later derived in the dual resonance model
\cite{Veneciano}.   
With the advent of the quark model of hadrons
and Quantum Chromodynamics, it became clear that
the fireball's temperature was in fact a transition
point to a new state of matter in heavy-ion
collisions, a plasma of deconfined quarks 
and gluons \cite{cabibbo}. 
On the other hand, a short-coming of the statistical
model was the isotropy in the decays being at odds of
the observed anisotropic angular distributions 
from cosmic ray
collisions. To explain this discrepancy, 
a two-fireball model was put forward, where both fireballs
could move according to an empirical velocity distribution to be 
determined from experimental fits.

Further developments arose in the sixties extending the
original Yukawa's idea on the exchange of
pions among nucleons to further explain the strong
interaction between hadrons in scattering
processes. In accordance with the copious pionization and 
dominance of short-range correlations, 
the multiperipheral model \cite{Frazer:1972xs}
of hadron inelastic collisions assumed that 
virtual hadronic objects exchanged between colliding hadrons
(pions, Regge trajectories, Pomerons...)
could themselves radiate
groups of correlated secondaries providing a theoretical
motivation for clusters of particles 
\cite{Meunier:1974nj}. 
Clustering of final-state particles
has been indeed experimentally confirmed in many
multi-particle processes  
but cannot be entirely identified with low-mass
resonances ($\rho$, $K^*$,...) (see e.g. \cite{Alver:2007wy,Ansorge:1988fg}).

It was quickly recognized, moreover, that the decays of two
or more clusters in an event could easily overlap
in (pseudo)rapidity \footnote{The longitudinal rapidity of a 
particle is defined as 
$y=(1/2)\cdot\ln{[(E+p_{\parallel})/(E-p_{\parallel})]}$
where $E$ and $p_{\parallel}$ denotes respectively
the energy and longitudinal momentum along the beam direction. 
Experimentally it becomes customary to measure a closely
related variable, the pseudorapidity, defined as
$\eta=(1/2)\cdot\ln{[(p+p_{\parallel})/(p-p_{\parallel}]}$
where $p$ denotes the particle momentum;  
rapidity and pseudorapidity coincide as far as $E \simeq p$,
i.e. when the particle mass is negligible as compared
to the transverse momentum. This is generally the case
for most secondaries (mostly pions) in a high-energy
inelastic collision. From an 
experimental point of view, the pseudorapidity
can be obtained directly from the measurement
of the polar angle $\theta$ of the emitted particle with respect
to the beam direction: $\eta=-\ln{(\tan{\theta}/2)}$.}, 
rendering the assignment of particles 
to one or another cluster an almost impossible task. Actually,  
the question arises whether such clusters are
real hadronic objects or purely statistical
artefacts to describe probability distributions
for high-energy multiparticle collisions.

Recently, following the Feynman gas picture \cite{Wilson,Andersson:1998ec}, 
a semi-empirical formula based on the formation and
decay of two $\lq\lq$cylinders'' has been formulated to
account for the (pseudo)rapidity distributions of charged
particles in hadronic inelastic collisions \cite{Liu:2008zzm}. 
At high enough energy,
both projectile and target cylinders are expected to overlap
only partially, developing a rapidity gap between both
fragmentation regions. Linear relationships
between the width of charged (pseudo)rapidity density distributions
and $\ln{\sqrt{s}}$ are found, in good agreement with experimental
data for p-p and nucleus-nucleus (A-A) 
collisions. Moreover, 
multiplicity distributions 
in proton and heavy-ion collisions 
can be included in a unified description where particles
are emitted from clusters formed at the collision \cite{Liu:2008zz}.

\subsection{Inclusive analysis: limiting fragmentation and plateau rise}

Investigations of multiparticle production basically rely upon
experimental results on:
total cross sections, multiplicity distributions, transverse
momentum and single-particle inclusive rapidity distributions,
two-, three-, or many-particle correlations, azimuthal correlations,
etc. Because of the huge amount of particles produced
in hadronic collisions at high energy, and the obvious experimental
difficulty for complete detection, inclusive analyses of events 
prove indeed to be extremely useful.

As usual, by one-particle (two-particle)
inclusive process we understand a reaction such that 
\beq\label{inclusive}
a\ +\ b\ \to\ c\ +\ (d)\ +\ X
\eeq
where $c$ ($d$) denote(s) here one (two) charged particle(s) and $X$
is an unknown system of particles. At LHC energies most
of the emitted charged particles will be pions.

Feynman in 1969 gave a description of hadronic interactions 
leading to the conclusion that as $s \to \infty$ 
($\sqrt{s}$ is the center-of-mass energy)
the inclusive cross section, expressed in terms
of $p_{\bot}^2$ and $x=2p^*_{\parallel}/\sqrt{s}$,
where $p^*_{\parallel}$ ($p_{\bot}$)is the momentum parallel (transverse)  
to the incident direction of the outgoing particle $c$ in the 
center-of-mass system (c.m.s.)
of the collision, becomes independent of the energy:
\beq
E\frac{d\sigma_{in}}{dp_{\parallel}dp_{\bot}^2}\ \to\ f(p_{\bot},x)
\eeq

More or less at the same time, Yang and collaborators \cite{Benecke:1969sh}
from a rather different point of view suggested 
the hypothesis of {\em limiting fragmentation}
whereby at high enough energies the inclusive cross section
for the production of particle $c$ from a target (or 
projectile) should be independent of the incident
energy and type of the other collision partner \cite{Otterlund:1978bw}.
For $x$ away from $x=0$, these two kinds of
scaling are equivalent, while at $x=0$ Feynman
scaling implies that the inclusive rapidity distribution
develops a {\em constant} (i.e. $\sqrt{s}$-independent)
plateau (see Fig.1).

\begin{figure}[ht!]
\begin{center}
\includegraphics[scale=0.58]{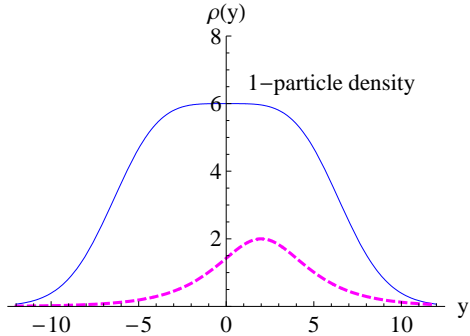}

\caption{Idealized single-particle inclusive spectrum $\rho_1(y)$
as function of the c.m.s. rapidity
developing a plateau in the central region.
The normalization is such that the area under the curve
represents the average number of charged particles $\langle n \rangle$
in p-p collisions at the LHC.
The dashed curve corresponds to one-particle density $\rho_1^{(s)}(y)$
from the (isotropic) decay
of a single resonance, cluster or fireball of mass $M \simeq 100$ GeV
moving with rapidity $=2$. 
The  available rapidity interval for the cluster decay
should grow as $\ln{M^2}$. 
We will make later use of this fact
in our study on possible long-distance correlations
stemming from a HS. This figure is intended for illustration
purpose only.}
\label{fig:plotrap}
\end{center}
\end{figure}

When expressed in terms of the rapidity variable, limiting
fragmentation requires the {\em correlation length hypothesis}
which states that there is no correlation between particles
whose rapidities $y_i$ differ by more than a certain correlation
length $\xi_y$ \footnote{There is no compelling
reason for a unique correlation length but we
will use one symbol for simplicity.}
; that is for $|y_i-y_j| >> \xi_y$. Moreover, there is
no correlation with the colliding particles as long as
their rapidities $y_a$ and $y_b$ differ from $y$ by a
distance large compared to $\xi_y$.
Therefore, considering a single-particle spectrum, whose maximum length
\footnote{The total length
of the rapidity plot is sometimes defined by the rapidities of the 
colliding particles $Y=|y_a-y_b|$, sometimes as $Y=\ln{(s/m_{\pi}^2)}$
where $m_{\pi}$ is the pion mass. The
latter choice means that $y_a$ and $y_b$ are not exactly at the ends
of the rapidity plots. In practice, the difference is insubstantial.} 
will be denoted by $Y$, three regions can be
distinguished \cite{Frazer:1972xs} in the c.m.s.:
{\em a)} {\bf Target fragmentation region}, where
$y^*<Y/2-\xi_y$; {\em b)} {\bf Central region}, where $|y^*| \lesssim 
Y/2-\xi_y$
and the aforementioned plateau rises; {\em c)} 
{\bf Projectile fragmentation region}, where $y^*>Y/2-\xi_y$.

Obviously at hadron colliders there is a symmetrical situation regarding
both target and projectile regions, potentially useful for
determining systematic uncertainties in inclusive analyses.
At LHC energies $Y \gtrsim 20$; however,  
we will implicitly restrict our study throughout this paper
to the central region $| y^* | \lesssim 2.5$. Notice that 
a distorsion (dip) is introduced in
the central plateau whenever  
the pseudorapidity variable is used in lieu
of the rapidity.

The early and naive expectation that the rapidity density (i.e. the
height of the plateau)
would be independent of $\sqrt{s}$, and that the plateau would merely
widen as the energy is increased, leading
to a logarithmic growth of multiplicity \cite{Wilson},
is clearly not in accord with data.
It rather turns out that the total (charged) multiplicity  
grows as $\langle n \rangle =a_2\ln{}^2\sqrt{s}+a_1\ln{\sqrt{s}}+a_0$ 
\cite{Busza:2007ke} (see \cite{Sarkisyan:2005rt,Akindinov:2007rr}
for a discussion on the {\em dissipating/effective} energy 
in the collision). 
This fact can be seen as a phenomenological consequence 
of three concurrent features: 
{\em a)} The increase as $\ln{\sqrt{s}}$ of the particle density
at midrapidity, {\em b)} The almost trapezoidal shape 
(see Fig.1) of the
(pseudo)rapidity distribution, and {\em c)} the growth of the
width of the spectrum proportional to $Y \sim \ln{\sqrt{s}}$.
Let us note that 
the predicted density in the central region at
LHC energies lies between 6 and 8 charged particles
per rapidity unit, depending on the MC generator 
and model \cite{Moraes:2007rq,Mitrovski:2008hb}.

Experimental data indeed have shown that Feynman scaling 
does not hold for the low $| x |$
(central) region, but limiting fragmentation does.
In fact, it turns out that limiting fragmentation
extends to a much wider range of rapidities than
originally thought, while 
the width of the central plateau grows
much slower than initially anticipated. 
According to Ref.\cite{Bialas:2004kt}, 
these features can be understood
from a two-step process in hadron collisions where 
the key-point is the flat distributions (in rapidity)
of both radiating partons and clusters in the string decay.

Still the dynamical origin of the
rise with energy of the central plateau is not
known though different models can account for it.
For example, it has been argued in \cite{Capella:1992yb} about two
possible causes, namely: {\em a)} increasing overlap in rapidity
space of the fragmenting valence-quark chains, {\em b)} contribution
from additional chains.  
This interpretation is supported by the experimental observation
that the plateau rise is mainly originated from the small-$p_{\bot}^2$
region. Finally, let us point out that $\langle p_{\bot} \rangle$
is remarkably independent of $\sqrt{s}$ for low multiplicities,
while exhibits a considerable increase for high multiplicity events
(see \cite{Kittel:2004xr,Kittel:2005fu} for a general review
on hadron collisions with references therein).

\subsection{Multiplicity distributions}

Another important result to understand multibody dynamics is
the shape of the (charged) multiplicity distribution, $P_n$. 
In 1972 Koba, Nielson and Olesen \cite{Koba:1972ng} 
proposed that the function 
\beq\label{eq:KNO}
P_n \cdot \langle n \rangle \equiv \psi(n/\langle n \rangle)
=\frac{\langle n \rangle \sigma_n}{\sigma_{in}} 
\eeq
should become asymptotically independent of $\sqrt{s}$, 
where $\langle n \rangle$ is the average number of 
(charged) particles per event, 
$\sigma_n$ and $\sigma_{in}$ denote the cross section
for production of $n$ charged particles and total
inelastic cross section, respectively.
In other words,
according to the KNO hypothesis the multiplicity
distribution for a given type of reaction
depends only on the ratio of the
number of particles to the average multiplicity.
Note that exact KNO scaling implies 
that particle correlations in general, and the normalized
moments of the multiplicity distribution in particular: 
$C_q=\langle n^q \rangle/\langle n \rangle^q$, 
are independent of $\sqrt{s}$
in contradiction with experimental observation 
\cite{Kanki:1989cx,Kittel:2005fu}.

In fact KNO scaling was approximately observed up to ISR energies
in p-p collisions. However, as the c.m.s. energy increases,
the multiplicity distribution becomes broader, developing
a $\lq\lq$shoulder'' structure at high $n$, hence 
breaking KNO scaling \cite{Bouzas:1992gg}.
Such a shoulder is usually ascribed to the multi-source
nature of the processes, either
from additional ladders, multi-parton interactions, or the presence of 
two components in events: one related to
the conventional soft interaction,
the other to QCD minijets. 
It is relevant to point out here
a generalization of KNO-scaling (log-KNO) 
\cite{Hegyi:1999aa} (based on Polyakov's original
idea \cite{Polyakov:1970xd})
which naturally arises for self-similar multiplicative
cascades. Any possible breaking of log-KNO scaling at the LHC
would be a hint of a different behaviour of the parton cascade,
deserving attention in our proposal.

On the other hand, the negative binomial 
distribution (NBD) has been extensively used
to describe empirically the multiplicity
distributions in almost all inelastic, high energy processes.
It is given by
\begin{equation}\label{NBD}
P_n=\frac{\Gamma(n+k)}{\Gamma(n+1)\Gamma(k)}
\biggl(\frac{\langle n \rangle}{k}\biggr)^n
\biggl(1+\frac{\langle n \rangle}{k}\biggr)^{-n-k}
\end{equation}
where $k$ is an adjustable parameter 
describing the shape of the distribution \cite{Dremin:2000ep}; 
the variance can be written as:
${D^2}/\langle n \rangle^2= \langle n \rangle+1/k$.
Thus, the NBD is broader than the Poisson distribution provided that
$k$ is positive and finite (corresponding to
positive correlations among particles), 
becoming Poissonian when $k \to \infty$.
The Bose-Einstein distribution is a special case of NBD with $k=1$.
If $k$ is negative, the NBD becomes a positive binomial distribution,
narrower than Poisson (corresponding to negative correlations).

Actually, the NBD does not provide a good description of particle correlations 
but a of convolution of different NBD's are 
required to fit experimental data \cite{Dremin:2004ts,Sarkisian:2000ux}.
Nonetheless, the wide occurrence of the NBD in different types
of high-energy collisions, suggested the interpretation
of such distribution law in terms
of {\em clans} \cite{Giovannini:1985mz}
(see \cite{Giovannini:2004yk} for a review). The clan concept 
generalizes the jet concept: 
all multiplicity correlations among particle
are exhausted within each clan. The new proposed parameters
are the average number of groups of particles 
with a common ancestor denoted here
as $\langle N_c \rangle$, and the average
number of particles per clan, $\langle n_c \rangle$.
These parameters are related to the standard ones
by: $\langle N_c \rangle =k\ln{(1+\langle n \rangle/k)}$
and $\langle n_c \rangle=\langle n \rangle/\langle N_c \rangle$,
where $1/k$ can be interpreted as a 
determination of the aggregation of partons into clusters/clans. 
Recently \cite{Brambilla:2006zt}, the parameter $k$ has been related to
a phase transition at parton level implying a discontinuity
in the average number of hadrons in high-energy hadronic collisions.

Finally, it is particularly interesting for us to note that the
average number of clans has been found to be decreasing with
$\sqrt{s}$ in semihard and hard processes (even approaching one
in a certain class of events \cite{Giovannini:2003ft}) 
while the mean number of particles per clan
should increase accordingly \cite{Brambilla:2006zt,Giovannini:2001kj}.

\subsection{Hadron-hadron versus $e^+e^-$ collisions}

Multiparticle production 
in hadron interactions might look at first glance
as completely different from $e^+e^-$ initiated processes. In the
latter case, one expects at tree level highly virtual timelike
partons. On the contrary, one generally describes a
hadron-initiated process 
as ladders of multiperipheral-type graphs between participating
constituents with low space-like virtualities. 
Nevertheless, a more unified picture emerges when one considers that
strings stretched between colour charges may fragment yielding
final particles through an intermediate clusterization stage
(see for example \cite{Kaidalov:1983ew,Andersson:1983ia}).

In fact, one striking feature of multiparticle production is simplicity
and universality \cite{Busza:2007ke,Kittel:2005fu,Back:2003xk}. For example, 
$\lq\lq$nuclear effects''
in h-A high-energy collisions are not really nuclear but generated by the 
QCD partonic medium to a large extent \cite{Cunqueiro:2008uu}. Indeed 
p-p, p-A and A-A collisions were studied in \cite{Sarkisyan:2005rt} 
providing a universal production mechanism based on 
$\lq\lq$dissipating energy participants'' in a Landau
hydrodynamical model \cite{Belenkij:1956cd}. 
Nevertheless, similarity between different
types of high-energy collisions may be more qualitative
than quantitative as stressed in \cite{Dremin:2004zy}.

Current models of high-energy hadron collisions combine
both perturbative QCD whenever applicable (high $p_{\bot}$
scattering) and a phenomenological approach to describe
soft physics. 
Frequently employed, local parton-hadron duality implies
the proportionality of inclusive distributions, becoming
asymptotically correct at ultrahigh energy, thus expectedly
at the LHC. (Notice however possible problems at low $p_{\bot}$ 
\cite{Abbiendi:2006qr}.) Actually, the subdivision of inelastic
hadronic collisions into soft and hard processes is
rather artificial, with semi-hard processes as a clear 
argument. In fact it is difficult to find a model
with the right mixture of both components.
In general, soft production processes are so unavoidably complex
that a sophisticated approach is 
definitely required \cite{Sjostrand:1987su,Sjostrand:2008vc}
implemented in Monte Carlo generators.

In the Dual Parton Model \cite{Capella:1992yb}
soft and semi-hard particle production in p-p 
inelastic collisions are understood
as a multiple Pomeron exchange between interacting
partons giving rise to sea-quark multi-chains 
which in turn fragment into final-state hadrons. 
According to a widely accepted picture, colour-strings
are stretched between participating partons, acting as
sources of particles successively broken by
the creation of $q-\bar{q}$ pairs from the sea.
The most common parton subcollisions are $gg \to gg$ 
subsequently yielding two strings between the hadrons
remnants. Initial state radiation can give extra gluons
contributing to the activity of the event.

The Lund model \cite{Lund} deals with string fragmentation
and cascades based on linear QCD confinement of quarks 
as the starting point and colour field breaking-up. 
Several Monte Carlo generators
(PYTHIA, FRITIOF, etc) implement this model
with further assumptions for hadronization at
the final stage.  
In PYTHIA \cite{Sjostrand:2007gs}, the perturbative approach 
expected to be valid at high-$p_{\bot}$ is extended down to
the low-$p_{\bot}$ regime, where a $p_{\bot min}$ cutoff
is required. Decreasing $p_{\bot}$ implies increasing
the number of parton-parton interactions, ultimately leading to
unitarity violation. A picture
based on multiparton interaction (let us recall
KNO breakdown) has been proposed to solve this problem.

On the other hand, 
assuming that pairwise parton interactions take place independently
of each other, the number of interactions
(thereby the number of strings/clusters emerging
from them) in an event
should obey a Poissonian distribution. However, since
hadrons are also extended objects, collisions may vary
over a range of impact parameters from very central to rather
peripheral ones, yielding extra correlations among secondaries. 
One net effect is a widening of the Poissonian distribution.

In sum, all the above-mentioned frameworks 
(see also \cite{Muller:2000zf,Arakelyan:2007kq} for other approaches)
cope with the complexity of hadron dynamics by invoking in essence
a two-step scenario; the resulting multiplicity distribution
is given by the convolution of the distribution for
particle emission sources (strings, clusters/clans, fireballs or whatever) 
with the decay/fragmentation distribution of the sources. 
This is basically
the approach followed in this work considering 
a prior stage in the cascade stemming
from a HS, leading to a {\em three-step
scenario provided that the parton-parton collision is hard enough}.

\subsection{Quark-Gluon Plasma}

In ultra-relativistic heavy-ion collisions, the energies of
the colliding particles are converted into thermal energy.
A hot state of matter, known as quark gluon plasma (QGP), 
should emerge if the energy density is large enough. As the 
plasma expands it cools down, undergoing a phase transition 
from deconfined QGP to confined hadrons. Strongly interacting
dense matter has been created in A-A
collisions at Brookhaven RHIC \cite{Adcox:2004mh} and will be
soon further studied at CERN ALICE experiment \cite{alice}.

A traditional signal sensitive to
deconfinement has been the expected suppression of $J/\psi$ production 
in nuclear reactions \cite{Matsui:1986dk,Brambilla:2004wf}, 
compared to the rates extrapolated from p-p data.
More generally, different signatures for QGP can be roughly divided
into two groups \cite{Arsene:2004fa}: 

\begin{itemize}
\item Modifications of specific properties of particles.
RHIC has in fact provided evidence for the creation
of a new state of thermalized matter exhibiting
almost ideal hydrodynamical behaviour based, among
others, on the following facts in addition to charmonium
suppression: strong jet quenching 
(implying an important parton energy loss in the medium ), 
strong elliptic flow (a signal of early thermalization
and a very low viscosity of the medium), direct photon
emission... 

\item Consistency of bulk properties with QGP formation, e.g.
entropy growth and behaviour of thermodynamic variables, 
correlations and fluctuations of the particle number 
\cite{Amelin:1994mf,Antoniou:1998ud,Kovchegov:1999ep,Voloshin:2005qj}, 
etc. Note that the correlation lengths can be good
indicators for a phase transition. 
Results from two-particle correlations observed at RHIC
\cite{Adams:2006yt} have recently been extrapolated to the LHC 
energy regime
with a special interest in jet studies \cite{Noferini:2007zz}.

\end{itemize}

Even though the postulated extra stage from a HS  
in the parton shower of p-p collisions is conceptually completely 
different from QGP, we will still find suggestive analogies
between both in our study of dynamical
fluctuations and correlations in particle emission. 
In fact, long-distance effects in (pseudo)rapidity
could become a distinctive signature of a non-standard 
parton cascade, as later argued. Nevertheless,
studies based on a description of inelastic hadronic 
interactions based on the Color Glass Condensate (CGC)
\cite{McLerran:1993ka}
also predict long-distance correlations 
\cite{Armesto:2006bv,Kovchegov:1999ep}.
It is thus crucial to discriminate between those
different sources of long-distance correlation effects
if NP effects were claimed to be observed in this way.

\subsection{Glasma}

Understanding thermalization of the medium 
in heavy-ion collisions yielding QGP is a still open
problem in the field as 
stressed in \cite{Kovchegov:2008qn}. 
According to the picture presented in 
\cite{McLerran:2008es}, almost instantaneously 
after the 
collision an ensemble of flux tubes
of rapidity-independent
color electric and magnetic fields (the Glasma)
is generated, decaying into quarks and gluons
eventually evolving into QGP.  
Particles would be produced
isotropically and with equal probability 
along the length of such Glasma tubes
spanning over large distances in rapidity, of order 
$1/\alpha_s(Q_{sat})$, where $Q_{sat}$ denotes the momentum
saturation of partons in the nucleon wavefunction.
Notice that hadronization models inspired in string fragmentation
(mentioned in section 2.4 regarding the Lund model)
also provide LRC due to impact parameter fluctuations. However, 
one peculiar new feature of Glasma is related 
to fields localized in the transverse scale over 
distances $1/Q_{sat}$ smaller than the nucleon size.

Experimentally, STAR has found LRC in heavy-ion collisions at RHIC
\cite{Srivastava:2007aw}, 
especially at large centrality (implying a larger number
of participants), likely larger
than can be expected from impact parameter
fluctuations. Plotting the correlation strength
versus the particles azimuthal angle, a structure
elongated in rapidity and collimated in angle appears
which can be interpreted in the Glasma framework \cite{Dumitru:2008wn}. 
Let us remark that these features of the ridge events are not seen in
proton-proton or deuteron-gold interactions and
seem to be unique to nucleus-nucleus collisions.

\subsection{Minimum bias, underlying events and jets at the LHC}

It is well known that high energy hadronic interactions are
dominated by the production of a large number of particles
with small transverse momentum $p_{\bot}$, leading to
the apparently striking conclusion that such 
collisions, governed by the strong-interaction, 
generally are rather $\lq\lq$soft''. 
As in all previous hadron colliders,
soft interactions will still be the dominant 
processes at the LHC, and a lot of potentially
interesting information to be obtained from them 
on hadron dynamics and new phenomena
should not be overlooked. Notice, in particular,
that as most of the produced particles 
will have low transverse momenta, dynamical
correlations should show up in the longitudinal
phase space.

A minimum bias event (MBE) corresponds 
ideally to a totally inclusive trigger and
can be basically identified with a non-diffractive
interaction.
In practice, experiments at colliders define their own
minimum bias triggers which, however, are not
in essence too different among themselves. 
MBE's are dominated by soft interactions, although
there is also some contribution from (semi)hard 
parton scattering often yielding
(mini)jets in the final-state. 

Experimentally, data show a strong correlation
among the charged particles indicating $\lq\lq$jet structure''
in MBE's even for momenta as low as 1 GeV. 
Let us stress, however, that 
it is not straightforward to extrapolate what has been learned 
about MBE's, e.g., at the Tevatron to the LHC, though
one could expect MBE to have harder scattering at higher energies
\cite{Field:2007zz}.

Even though not completely clean from a theoretical viewpoint,
a third category of events is generally reported in
high-energy hadron collisions:   
semi-hard events. A semi-hard event, roughly speaking,
is characterized by the presence of minijets and
does not plainly fall into the soft and hard categories.
Semi-hard events are likely called to
play a role of utmost importance in our proposal.

On the other hand, the concept of underlying
event (UE) in a hard scattering process 
has become nowadays an important issue in 
collider physics \cite{Moraes:2007rq}. 
In a few words, an UE is 
everything but the outgoing hard jets, i.e. 
the event activity remaining once the jets have been removed.
As hard collisions are correlated, UE and MBE do {\em not}
coincide. 
In fact, an UE contains both soft and
hard interactions: the former mainly
corresponds to the beam-beam remnant interactions, 
while the hard component should be ascribed to the initial
and final-state radiation, from colour strings
stretching between the UE and the jets and from
secondary parton interactions.

Usually, it has been thought of collecting MBE's as important for
the general study of general features of p-p interactions.
Let us summarize below some of the main points 
aimed by the study of minimum and underlying events:
\begin{itemize}
\item (Re)tuning of MC generators to describe hadronic
interactions \cite{Skands:2007zz}, allowing a better understanding of
systematic uncertainties. Multiparticle
production should also serve as an important tool
for the calibration and understanding of detectors,
establishing a solid basis for exclusive channels.

\item Better knowledge of multiparton interactions,
particularly the interplay between minimum bias
and underlying events, allowing a better 
understanding of the structure of the proton at low-$x$. 
This in turn will facilitate a better control of jet 
backgrounds to SM and BSM physics processes.

\item Reference study for heavy-ion collisions.

\end{itemize} 

Let us add tentatively a new item to the above list:
In our proposal, soft events with a (semi)hard signature
tagging NP (possibly a kind of {\em anomalous underlying event} 
as pointed out in \cite{Harnik:2008ax})
will play an important role in the search strategy for a HS showing up
in the parton cascade.

\section{Hidden sectors}

\subsection{Unparticle stuff}

Recently, Georgi \cite{Georgi:2007ek}
introduced the notion of unparticle stuff, ${\cal U}$, 
based on the possibility of a scale invariant
HS, with experimental signals radically different
from the SM sector. Conformal invariance dictates the form of 
the propagator,
thereby determining the virtual unparticle contributions.
The unparticle stuff ${\cal U}$ may appear as a non-integer
number of invisible particles, $d_{\cal U}$. 
Actually unparticle does not have a fixed invariant mass;
from a quantum field framework it can be seen to exhibit a 
continuous mass spectrum.
Upon measurement, however, unparticle would behave
as an ordinary particle with a definite (but arbitrary!) mass
\cite{Nikolic:2008ax}.

If unparticle stuff couples to the SM Higgs boson,
electroweak symmetry breaking should break 
conformal invariance too, providing an effective
mass in the unparticle propagator. Thus unparticle physics
could be probed for energies together with the discovery
of the Higgs boson \cite{Kikuchi,Sannino}, typically 
in the window between 10 GeV and 1 TeV, i.e. at LHC reach.
Moreover, interactions can effectively induce a mass gap for 
the unparticle sector \cite{Fox:2007sy} 
sharing common features with
Hidden Valley models to be examined next. 
Finally note that unparticle physics also provides 
dark matter candidates \cite{Kikuchi2,Sannino}
with astrophysical and cosmological 
implications \cite{Davoudiasl:2007jr}.

It is widely accepted that unparticle stuff 
could manifest itself basically in two kinds of processes:
\begin{itemize}
\item[{\em a)}] Virtual effects, e.g., the unparticle
propagator mediates processes like $q\bar{q} \to \ell^+\ell^-$
or $gg \to \ell^+\ell^-$, or can in loop processes,  
leading to new signals at colliders in precision measurements
\cite{He:2008xv}.
\item[{\em b)}] Real production of unparticle stuff, implying
a new phenomenology in collider physics. The most interesting
signature would be peculiar missing energy distributions 
because of the non-integral values of $d_{\cal U}$. However,
it has been argued \cite{Neubert:2007kh}
that such signals may be less striking
than originally advocated.  
Nevertheless, unparticles might travel a macroscopic distance before
decaying, leading to delayed signals and displaced vertices.
\end{itemize}

Moreover, in Ref.\cite{Feng:2008ae} unparticle self-interaction
is considered as mediating processes such as
$gg \to \cal{U} \to \cal{U} \cdots \cal{U}$,
i.e. with two or more unparticles in the final-state.
It is phenomenologically interesting to realize that,
conversely to SM channels such as $gg \to g \cdots g$, 
the creation of additional unparticles in the final-state
does not suppress the rate of multi-unparticle production, for
strongly-coupled conformal sectors. This feature
is especially relevant for our proposal because of 
the fluctuations of the number of intermediate states 
in the hadronization process, though the requirement
of converting back to SM particles may imply
that they are sub-dominant.

As previously commented, missing energy and momentum
carried away by the unparticle have been claimed
as typical signatures, pretty much as for Kaluza-Klein modes
in a large extra dimension scenario. Several processes 
have been then envisaged based on real emission of
unparticle stuff: $t \to b\ \cal{U}$, Higgs $\to \gamma\ \cal{U}$,
quarkonium $\to \gamma\ \cal{U}$, etc (see \cite{Cheung:2008xu}
for a minireview). 

However, unparticle couplings to other
SM fields may give rise to an effective decay width
and, if kinematically allowed, unparticle 
excitations can decay to SM particles \cite{Rajaraman:2008bc,Delgado:2008gj}.
Therefore, the unparticle stuff might not be necessarily characterized by
missing energy signals, drastically modifying
the discovery prospects so far  
considered, and motivating the use of correlation techniques
as advocated in this paper. 

Intuitively, an intermediate unparticle
state in the parton cascade, whose mass is not fixed but 
obeys a continuum 
spectrum, 
should yield large fluctuations in a event by event basis. Consequently
some charactistic features of the showering should appreciably vary. 
How to assess such possible degree of modification is
the main goal of this paper.   

\newpage

\subsection{Hidden Valleys}

Another scenario related to unparticle physics with 
similar phenomenology is the 
so-called $\lq\lq$Hidden Valley'' (HV) \cite{Strassler:2006im}, 
where the SM is accompanied by a HS of new particles
not been yet observed due typically to an
energetic barrier or a weak coupling to SM particles.
These models are characterized by low
mass $\lq\lq$valley'' ($v$) states and a new
dynamics in the HS governed by the 
confinement scale which introduces a mass gap
into the theory.

The effects of a Hidden Valley are determined 
to a large extent by
the nature of the mediators like $Z'$,
Higgs sectors, gravitinos from Randall-Sundrum or large
extra dimensions. If the mediator mass is at reach
of the LHC one should expect observable effects. Indeed,
the mediator would connect the interacting partons 
of the colliding protons with the $v$-quarks,
subsequently forming $v$-hadrons. In turn, the
$v$-hadrons would decay into quarks (igniting
a true QCD partonic cascade) and leptons provided
that the mass gap of the theory is not too small
and the resonances from the HS not too light.
In a region of parameter space of the model, $v$-hadrons 
($\sim O(10)$ GeV) would decay promptly back into 
SM particles \cite{Strassler:2006im}. 
We are focusing on this possibility
in this paper.

Events in p-p inelastic collisions 
with such $v$-hadrons are expected to 
be more spherical, with lower thrust and
more isolated leptons, than those from SM background 
processes. Production of very massive clusters would 
provide a key observable
to identify the signal \cite{Han:2007ae}.

On the other hand, the study of jets is the traditional way
for hunting such kind of NP. 
The invariant mass of a $v$-quark initiated jet
should be substantially larger than that of
a standard jet, and the associated multiplicity
too. As a consequence, more particles and more 
separated in rapidity space will become correlated.
As a side effect, the possibility of merging partons from two 
different jets coming from HV becomes larger than in standard jets.
All these combined effects may lead to 
a signal with larger SM backgrounds. Moreover,
many {\em soft jets} in events rather than 
well collimated high-$p_{\bot}$ jets can become
a distinctive signature from a HV sector.

\subsection{Other scenarios}

Inclusive correlation techniques might be useful
to unravel a signal from a HS in other scenarios too. 
First, Higgs bosons can be viewed as portals to the discovery
of a HS, as emphasized in \cite{Patt:2006fw}. For instance,  
non-standard Higgs bosons
or unstable neutralinos could decay into SM particles
through cascade processes \cite{Chang:2008cw}, 
notably under the hypothesis of 
the existence of light pseudoscalar Higgs bosons 
still evading current B-physics and LEP bounds \cite{Domingo:2008rr}.

Another interesting example is the Walking Technicolor model 
\cite{Belyaev:2008yj}, as heavy states like composite vector
resonances and Higgs could
show up in the hadronic cascade yielding final-state SM particles. 

Let us also mention the framework recently proposed in
 \cite{Kang:2008ea} where {\em quirks} emerge from an additional
unbroken non-abelian gauge group with some new fermions in the fundamental
representation. Long and stable
strings stretched between quirks would lead  
to a highly exotic phenomenology, depending crucially
on the length of the strings (which might be $\lq\lq$macroscopic'').
Interestingly, {\em quirkonium} states can be formed 
in p-p collisions at the LHC,
undergoing a prompt annihilation 
mainly via strong decay modes into gluons and quarks
for colourful quirks. 
The inclusive analysis advocated in this work can be appropriate
to deal with quirk phenomenology, e.g. when very 
massive states (hadronic fireballs?)
decay into light hadrons with $\cal{O}$(GeV) energy in  
very high multiplicity events, together with hard decay products
to be used as tags of NP.

Last but not least, mini black holes in large extra dimensions
theories are another potential source of
high-multiplicity events at the LHC \cite{Giddings:2001bu} 
where our approach might be helpful. Note
that the Hawking temperature wavelength should be larger
than the size of the black hole, implying
a point-like radiator behaviour mainly emitting
$s$-waves.

\newpage

\section{Inclusive Correlations}

In anticipation of the difficulty of distinguishing  
between different HS scenarios at the LHC,  
inclusive correlation techniques between emitted
particles could provide a well-tested
tool supplying complementary information 
on the possible formation of intermediate
states in the parton cascade and their nature. To this aim, we
examine below 2-particle correlation functions.

\subsection{Two-particle rapidity correlations}

A general inclusive 2-particle correlation function 
for inelastic collisions can be defined as 
\cite{Foa:1975eu}
\begin{equation}\label{eq:corgen}
C_2(y_1,p_{\bot 1},y_2,p_{\bot 2})= 
\frac{1}{\sigma_{in}}\frac{d^6\sigma}{dy_1d^2p_{\bot 1}dy_2d^2p_{\bot 2}}
- \frac{1}{\sigma_{in}^2} \frac{d^3\sigma}{dy_1d^2p_{\bot 1}}
\frac{d^3\sigma}{dy_2d^2p_{\bot 2}}
\end{equation}
where $\sigma_{in}$ denotes the inelastic cross section and
the subscripts 1 and 2 refer to the two considered particles, 
event by event. For the sake of simplicity we will not 
distinguish between different species of particles, focusing
only on charged particles.

As already mentioned in the Introduction, we will
confine our analysis to (longitudinal) rapidity
correlations in p-p interactions
\footnote{There are studies \cite{Porter:2005gp}
on transverse rapidity correlations in p-p collisions
not considered here.}. 
Therefore, we perform an integration of (\ref{eq:corgen}) 
over the whole $p_{\bot}$ range, getting a new
$p_{\bot}$-independent 2-particle correlation function
\begin{equation}\label{eq:C2}
C_2(y_1,y_2)= 
\frac{1}{\sigma_{in}}\frac{d^2\sigma}{dy_1dy_2}
- \frac{1}{\sigma_{in}^2} \frac{d\sigma}{dy_1}
\frac{d\sigma}{dy_2} \equiv \rho_2(y_1,y_2)-\rho_1(y_1)\rho_1(y_2)
\label{eq:corfunctionrap}
\end{equation}
where we have introduced the single and double
rapidity densities of charged particles:
\begin{eqnarray}\label{eq:definitions}
\rho_1(y) &=& \frac{1}{\sigma_{in}}\ 
\int d^2p_{\bot}\frac{d^3\sigma}{dyd^2p_{\bot}} \\
\rho_2(y_1,y_2) &=& \frac{1}{\sigma_{in}}\ 
\int d^2p_{\bot 1}d^2p_{\bot 2}\ 
\frac{d^6\sigma}{dy_1d^2p_{\bot 1}dy_2d^2p_{\bot 2}}
\end{eqnarray}
with the normalizations obtained by integration over again the whole
rapidity range $Y$
\begin{equation}{\label{eq:norm}}
\int\ dy\ \rho_1(y)=  
\langle n \rangle\ ;\ 
\int\  dy_1\ dy_2\  \rho_2(y_1,y_2) =  
\langle n(n-1) \rangle\ ;\ 
\int\ dy_1\ dy_2\ C_2(y_1,y_2)= 
\ D^2-\langle n \rangle
\end{equation}
where $D^2=\langle n^2 \rangle - \langle n \rangle^2$
is the variance of the charged emitted particles.

It is customary to define a related correlation function
less sensitive to experimental errors as
\beq\label{eq:R2}
R_2(y_1,y_2)= \frac{C_2(y_1,y_2)}{\rho_1(y_1)\rho_1(y_2)}=
\frac{\rho_2(y_1,y_2)}{\rho_1(y_1)\rho_1(y_2)}-1
\eeq

A very instructive description
of the behaviour of $R_2(y_1,y_2)$ is provided by
a two-dimensional plot in which contour-lines
for $R_2$ values are shown 
for p-p collisions \cite{Foa:1975eu}; similar
analyses have been performed for proton-emulsion interactions
\cite{Simic:1979pp}).
If one cuts the dimensional plot
by planes along lines of fixed $y_1$ it is generally
found a peculiar triangular
shape pointing out the existence of short-range correlations
(SRC) \cite{Foa:1975eu,Simic:1979pp,Breakstone:1982hg}, 
of order of one unit of rapidity. We will turn to these
issues in sections 5.3 and 5.4.

\subsection{Short- and long-range rapidity correlations}

As commented in the Introduction, SRC are ubiquitous in
multiparticle hadroproduction.
Motivated by the Feynman fluid analogy \cite{Wilson}, one can write
\beq\label{eq:C2S}
C_2^{SR}(y_1,y_2) \sim \exp{[-|y_1-y_2|/\xi_y]}
\eeq
where $\xi_y$ denotes the correlation length. Typically
$\xi_y$ has been found to be of the order of one rapidity unit
in hadron inelastic collisions \cite{Foa:1975eu}.

A Gaussian shape, characteristic for particle production 
from clusters, is often employed:
\beq\label{eq:C2Gauss}
C_2^{SR}(y_1,y_2) \sim \exp{[-(y_1-y_2)^2/4 \delta^2]}
\eeq
Notice that, in order to obtain the same average $|y_1-y_2|$,
$\delta$ and $\xi_y$ are related through:
$\delta=(\sqrt{\pi}/2) \cdot \xi_y$.

Even if particle emission from the created sources were totally
uncorrelated, long-range correlations (LRC) would occur in the process
of averaging over fluctuations in the number of emitting sources
as pointed out in \cite{Capella:1978rg}.
It is easy to see
that this picture actually implies LRC
in addition to SRC originated in the cluster decay.

More generally, the inclusive correlation function is
usually split  in two terms:
\begin{equation}{\label{eq:shortlongcorr}}
C_2(y_1,y_2)=C_2^{LR}(y_1,y_2)+C_2^{SR}(y_1,y_2)
\end{equation}
where the {\em short-range} term $C_2^{SR}$ is generally
assumed to be more sensitive to dynamical correlations, while
$C_2^{LR}$ stands for (somewhat misleadingly called) 
{\em long-range} correlations, usually
arising from mixing different topologies 
(i.e. from unitarity constraints \cite{Capella:1978rg,Ansorge:1988fg})
in events.

It is very important to point out that the rapidity range 
kinematically available for a decay particle
from a cluster/clan (or particle) of mass $M$ 
grows as $\ln{M^2}$. Thus it is perfectly plausible that
the correlation length $\xi_y$ itself gets increased 
following a similar logarithmic dependence. As $\xi_y$ has been
found to be about one rapidity unity in experiments
so far, we may reasonably speculate that in the presence of heavy
sources of clusters or particles, the rapidity correlation
length happens to be significantly larger $\xi_y \sim \ln{(M_u^2/M^2)}$
where $M_u$ can be typically of order 100 GeV (or more), while
$M$ can be taken of $\cal{O}$(GeV). 
{\em Therefore, the correlation length 
between produced (charged) particles, not coming from
unitarity constraints, could
reach values much larger than one rapidity unit.}
One should prevent however
a possible double counting since then some LRC effects 
could be then partially attributed to the $C_2^{SR}$
piece through a larger $\xi_y$ value, while 
the $C_2^{LR}$ piece resumes the LRC $\lq\lq$by definition''. 
In any event, the $C_2^{SR}$ term (to be identified later with cluster
decays) has to exhibit a correlation length longer
than expected in a conventional approach, if a massive stuff from
a HS is active in the cascade.

On the other hand, a 2-particle correlation function 
normalized to the mean particle density $\bar{\rho}_1$ 
in the central region can be parametrized as \cite{Abbott:1995as} 
\footnote{$R_2(y_1,y_2)$ and $\hat{R}_2(y_1,y_2)$
practically coincide in the central rapidity region
of both particles.}
\beq\label{eq:R2hat}
\hat{R}_2(y_1,y_2)=\frac{C_2(y_1,y_2)}{\hat{\rho}_1^2}=
\alpha\ e^{-|y_1-y_2|/\xi_y}\ +\ \beta
\eeq
Integration of this function 
over different (pseudo)rapidity windows 
has been carried out to extract $\alpha \cdot \xi_y$
from experimental data in heavy-ion collisions, 
to obtain information about
a possible phase transition in QGP formation 
\cite{Abbott:1995as}.

To end this part, let us briefly mention Bose-Einstein statistics leading to
specific positive and SRC 
between identical bosons. Bose-Einstein correlations (BEC)
are a well-established effect \cite{Csorgo:2004sr} 
in all types of collisions. 
However, the contribution of BEC
to intermittency of scaled moments is expected to be small 
when the rapidity variable is used \cite{Andreev:1995rc},  
and thereby will not be considered in this work.

\newpage

\subsection{Normalized factorial moments}

The study of fluctuations in particle physics 
has a long history \cite{Kittel:2005fu,DeWolf:1995pc} 
starting with early cosmic-ray observations
on large concentration of particle number in small rapidity regions.
The main point
is whether such effect is of a dynamical or merely
statistical origin (see \cite{Capella:1989xg,Abbott:1995as,
Abbiendi:2001bu,Tawfik:2000cn,Li:2007zzt}
for various production processes). 
Factorial moments of the multiplicity
distribution are the usual tool to try to give an answer
to this question, and will be employed in our analysis
seeking a HS in parton cascades.

First, let us briefly review the main definitions and properties
of some moments of the multiplicity distribution 
\cite{DeWolf:1995pc,Kittel:2005fu}.
The normalized moment of rank $q$ of a multiplicity
distribution $P_n$ is given by 
\begin{equation}\label{eq:Fq}
F_q=\frac{\sum_{n=q}^{\infty}n(n-1) \cdot\cdot\cdot (n-q+1)\ P_n}
{(\sum_{n=1}^{\infty}n\ P_n)^q}
\end{equation}
corresponding to the normalized phase-space integral over
the $q$-particle density function 
\footnote{Cumulant factorial moments $K_q$,  
representing the normalized phase-space integrals over the $q$-particle
correlation function, can be 
obtained from the factorial moments: $F_2=1+K_2$, $F_3=1+3K_2+K_3$, etc.
For a Poissonian distribution 
$K_q=0,\ \forall q$. Since $\mid K_q \mid$ and $F_q$ increase
quickly with increasing $q$, it is useful to define \cite{Dremin:1993fr}
the ratio $H_q=K_q/F_q$ which keeps the same order of magnitude
over a large range of $q$. Normalized cumulants are 
often associated to {\em genuine} correlation patterns 
in experimental data \cite{Abbiendi:2001bu}. 
For the sake of simplicity they are not  
employed in this work.}. 

The second order $F_2$ moment in particular, is 
related to the 2-particle correlation function as:
\begin{equation}\label{eq:F2}
F_2= \frac{\int\ dy_1dy_2\ C_2(y_1,y_2)}
{[\  1/\sigma_{in}\int\ dy\ d\sigma/dy\ ]^2}+1=
\frac{\langle n(n-1) \rangle}{\langle n \rangle^2}
\end{equation}
$F_2$ can be mathematically related to the
$k$ parameter for a NBD according to:
$F_2=1+1/k$, hence determining the degree of correlation
between the produced particles.

Higher $F_q$ moments can be
related to $F_2$ in the Ising model as \cite{Chau:1992uq}  
\begin{eqnarray}\label{eq:Fqrelations}
F_3 &=& \frac{3}{2}F_2^2-\frac{1}{2} \\
F_4 &=& 3F_2^3-3F_2+1 \\
F_5 &=& \frac{15}{2}F_2^4-15F_2^2+10F_2-\frac{3}{2}
\end{eqnarray}
This set of relations will play a role of paramount importance in
our toy simulation as we shall see, though other possible ansatzs 
(see for example \cite{Capella:1989xg})
would lead to the same final conclusions. 

On the other hand, the normalized factorial moments $F_q$ can be
defined both in full phase-space, as well
as in ever smaller intervals of it. If
the available rapidity space $\Delta Y$ is split into $M$ bins
(for simplicity) of equal size, $\delta y = \Delta Y/M$, 
the $F_q$ moment can be expressed as
\beq\label{eq:Fqbin}
F_q(\delta y) = \frac{\langle n_j (n_j-1) ... (n_j-q+1) \rangle}
{\langle n_j \rangle^q} 
\eeq
where $n_j$ 
denotes the multiplicity in the $j$-th bin, and 
$\langle n_j \rangle$ is the average multiplicity of the $j$-bins
in a single event and then averaged over all events
(the co-called vertical analysis at one bin).

The effect of growing factorial normalized 
moments with decreasing $\delta y$ ($\lq\lq$intermittency'')
was introduced by
Bialas and Peschanski \cite{Bialas:1985jb} 
in high-energy physics motivated by an analogy with 
turbulence in hydrodynamics, 
in order to describe enhanced fluctuations in the density
distributions. Intermittency can be understood as
a manifestation of SRC \cite{Carruthers:1989iw}, 
and is closely related to cascade models with scale invariant
branching structure, ultimately manifesting
a multifractal nature \cite{Carruthers:1983zx,Ochs:1988ky}.
We shall come back to this important issue in section 7.
Lastly, it is worthwhile noting 
that intermittency implies a sizable widening of the multiplicity
distribution in p-p collisions, together with a longer
tail at high mutiplicities.

\section{Cascades in hadronic collisions}

In this section we consider two- and three-step   
cascades yielding final-state SM particles in hadron-induced reactions.
We assume that events can 
be decomposed into groups of particles (generically called clusters or clans) 
according to a common ancestor, either parton, string, or even
an (un)particle from a HS. Furthermore, 
such clusters/clans/(un)particles are
treated as physically real objects in our toy model, 
with a production cross section $\sigma_c/\sigma_u$ in 
p-p inelastic collisions.
In neither case interference effects will be considered 
in our calculations.

\subsection{Two-step scenario}

Let us assume that final-state particles originate from
clusters formed in the first stage of an inelastic collision.
A tree-structure may of course happen, i.e. 
clusters may decay into clusters which finally lead to 
asymptotic particles, but production cross sections below
refer to $\lq\lq$primary'' clusters to which final-state particles
are associated.
  
Hence we introduce single and double
differential cross sections for cluster production  
$(1/\sigma_c)\ d\sigma_c/dy_c$ and $(1/\sigma_c)\ d^2\sigma_c/dy_{c1}y_{c2}$,
respectively, satisfying the normalizations
\begin{equation}{\label{eq:normclus1}}
\frac{1}{\sigma_c}\\ \int  dy_c\ \frac{d\sigma_c}{dy_c}\ =\ 
\langle N_c \rangle \ \ ;\ \  
\frac{1}{\sigma_c}\ \int dy_{c1}dy_{c2}\ \frac{d^2\sigma_c}{dy_{c1}dy_{c2}}\  
=\ \langle N_c(N_c-1) \rangle 
\end{equation} 
where $\langle N_c\rangle$ denotes the average number of produced
clusters per collision. All integrals in this section extend over the full
available rapidity interval in the collision.

Now, let $\rho_1^{(s)}(y,y_c)$ and $\rho_2^{(s)}(y_1,y_2,y_c)$
denote the one- two-particle
rapidity densities of (charged) hadrons emitted by a single
cluster with rapidity $y_c$, with normalization
\begin{equation}{\label{eq:rho1}}
\int dy\ \rho_1^{(s)}(y,y_c)\ =\ \langle n_c \rangle\ ;\ 
 \int dy_1dy_2\  \rho_2^{(s)}(y_1,y_2,y_c)\ = 
\langle n_c(n_c-1) \rangle
\end{equation} 
where $\langle n_c \rangle$ is the mean particle
multiplicity from a single cluster decay. 

In absence of any interference between the
different contributions, we can write the differential 
inelastic cross section as the convolution:
\begin{equation}{\label{eq:decay1}}
\frac{1}{\sigma_{in}}\frac{d\sigma}{dy}=
\frac{1}{\sigma_c}\ \int\ dy_c\ \frac{d\sigma_c}{dy_c}\ \rho_1^{(s)}(y,y_c)
\end{equation}
while for two emitted particles, we can write
\begin{equation}{\label{eq:decay2}}
\frac{1}{\sigma_{in}}\frac{d^2\sigma}{dy_1dy_2}
=\frac{1}{\sigma_c}\ \int\ dy_c\ \frac{d\sigma_c}{dy_c}\ 
\rho_2^{(s)}(y_1,y_2,y_c)\ 
+\ \frac{1}{\sigma_c}\ \int dy_{c1}dy_{c2}\ 
\frac{d^2\sigma_c}{dy_{c1}dy_{c2}}\ 
\rho_1^{(s)}(y_1,y_{c1})\ \rho_1^{(s)}(y_2,y_{c2})
\end{equation}
where the two particles may come either from a single cluster 
(first term on the r.h.s.) or from two
different clusters (second term on the r.h.s.). 

Accepting that clusters can be produced in a
correlated way \cite{Uhlig:1977dc}, let us introduce the 2-cluster
correlation function satisfying the usual normalization condition (right)
\begin{equation}{\label{eq:clustercorfunc}}
C_2^{(c)}(y_{c1},y_{c2})=\frac{1}{\sigma_c}\frac{d^2\sigma_c}{dy_{c1}dy_{c2}}
-\frac{1}{\sigma_c^2}\frac{d\sigma_c}{dy_{c1}}\frac{d\sigma_c}{dy_{c2}}\ \ \ 
;\ \ \  
\int dy_{c1}dy_{c2}\ C_2^{(c)}(y_{c1},y_{c2})=D_c^2 - \langle N_c \rangle 
\end{equation}
where $D_c^2=\langle N_c^2 \rangle - \langle N_c \rangle^2$ 
stands for the dispersion of the cluster distribution. 
Notice that for a Poissonian distribution of the cluster distribution
$D_c^2=\langle N_c \rangle$
or equivalently the integral (\ref{eq:clustercorfunc}) should be set
equal zero (valid for the often used Independent Cluster Emission model).

The result of combining the definition of the 2-particle correlation function
in Eq.(\ref{eq:C2}) with 
Eqs.(\ref{eq:decay1}-\ref{eq:decay2}), can be split into two pieces
following Eq.(\ref{eq:shortlongcorr}):
\begin{equation}\label{eq:C2sl}
C_2(y_1,y_2)
= C_2^{LR}(y_1,y_2)\ +\ \langle N_c \rangle\ C_2^{(single)}(y_1,y_2)  
\end{equation}
where the piece $C_2^{(single)}$ corresponding to 
2-particle correlations $\lq\lq$inside'' a {\em single} cluster, reads
\begin{equation}\label{eq:single}
C_2^{(single)}(y_1,y_2)=
\rho_2^{(s)}(y_1,y_2)-\rho_1^{(s)}(y_1) \rho_1^{(s)}(y_2) 
\end{equation}
Note that we have assumed above that $C_s^{(single)}(y_1,y_2)$ has no
explicit dependence on cluster rapidity, as one indeed 
expects an overall dependence on the $|y_1-y_2|$ difference
(see Eq.(\ref{eq:C2single}) below).

On the other hand, the $C_2^{LR}$ piece 
is ascribed (to be justified a posteriori) to the convolution
\begin{equation}\label{eq:convolution}
C_2^{LR}(y_1,y_2)=
\int dy_{c1}dy_{c2}\ D_2^{(c)}(y_{c1},y_{c2})\ \rho_1^{(s)}(y_1,y_{c1})\ 
\rho_1^{(s)}(y_2,y_{c2})
\end{equation}
where we have defined the function
\begin{equation}\label{eq:Dc}
D_2^{(c)}(y_{c1},y_{c2})=C_2^{(c)}(y_{c1},y_{c2})
+\frac{1}{\sigma_c}\frac{d\sigma_c}{dy_{c1}}\ \delta(y_{c1}-y_{c2})
\end{equation}
obeying the relation
\begin{equation}\label{eq:norm2}
\int dy_{c1}dy_{c2}\ D_2^{(c)}(y_{c1},y_{c2})\ =\ D_c^2
\end{equation}

Confining our attention to
a rapidity interval in the central
region (where single spectra are approximately constant)
we finally get 
\beq\label{eq:corre}
C_2(y_1,y_2)= (\bar{\rho}_1^{(s)})^2 \ D_c^2\ +\ \langle N_c \rangle\
C_2^{(single)}(y_1,y_2) 
\end{equation}
where $\bar{\rho}_1^{(s)}$ denotes the average charged particle density 
from a single cluster, obeying the relation
$\bar{\rho}_1 = \langle N_c \rangle \cdot \bar{\rho}_1^{(s)}$, 
where $\bar{\rho}_1$ denotes the average one-particle density
in p-p collisions, estimated to be 6-8 charged particles
per rapidity unit in the central plateau at LHC energies
\cite{Moraes:2007rq,Mitrovski:2008hb}.

From Eqs.(\ref{eq:C2sl}) and (\ref{eq:shortlongcorr}) 
one can tentatively identify the long-range
piece of the 2-particle correlation function as
$C_2^{LR}=(\bar{\rho}_1^{(s)})^2 D_c^2$, while
$C_2^{SR}= \langle\ N_c \rangle\ C_2^{(single)}(y_1,y_2)$
can be put in correspondence with the short-range piece if the correlation
length $\xi_y$ were actually small, as already anticipated.

Next, the factorial moment $F_2$ is obtained integrating 
(\ref{eq:corre}) over the allowed $(y_1,y_2)$ region, i.e.
\begin{equation}\label{eq:F2twostep}
F_2=F_2^{(c)}+\frac{F_2^{(single)}}{\langle N_c \rangle}
\end{equation}
where the scaled moment for clusters, $F_2^{(c)}$,
and particles inside clusters, $F_2^{(single)}$, are respectively given by
\begin{equation}\label{eq:F2single}
F_2^{(c)}=\frac{ D_c^2}{\langle N_c \rangle^2}-\frac{1}{\langle N_c \rangle}+1
=\frac{\langle N_c(N_c-1) \rangle}{\langle N_c \rangle^2}\ \ ;\ \ 
F_2^{(single)}=
\frac{\int\ dy_1dy_2\ C_2^{(single)}(y_1,y_2)}
{\langle n_c \rangle^2}+1
\end{equation}
We will assume later an exponential shape for $C_2^{(single)}(y_1,y_2)$, i.e.
\beq\label{eq:C2single}
C_2^{(single)}(y_1,y_2)=c_s\ e^{-|y_1-y_2|/\xi_y} 
\eeq
where $c_s \sim (\bar{\rho}_1^{(s)})^2$ and
$\xi_y \simeq 1$ in a conventional parton cascade. Explicit
integration of Eq.(\ref{eq:F2single}) will be carried out in section 7.

\begin{figure}[ht!]
\begin{center}
\includegraphics[scale=0.9]{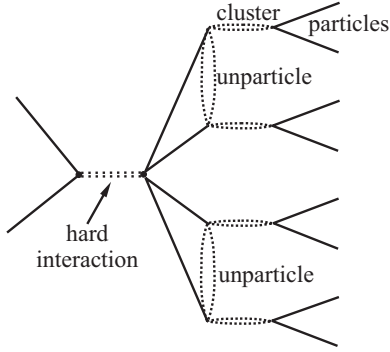}
\caption{Pictorial representation of a 3-step
scenario where unparticles (for this particular HS choice)
are produced in
a hard parton collision, subsequently decaying into
final-state SM particles through cluster formation.
The number of unparticles 
at the onset of the cascade may fluctuate,
and a large unparticle mass would induce additional 
long-distance correlations among the
final-state particles.}
\label{fig:drawing}
\end{center}
\end{figure}

\subsection{Three-step scenario}

Now let us consider a non-standard stage of matter 
(e.g. unparticle stuff \footnote{Our results
should remain essentially 
valid for the other possibilities mentioned in section 3.})
appearing from the primary (hard) parton interaction, on top of 
the conventional QCD parton cascade (see Fig.2)
\footnote{Even though the whole hadronization cascade
can be seen as a multi-step and self-similar process itself,   
the distinction between 2- and 3-step scenarios makes sense
as far as we are considering an extra non-standard stage.}. 
Such a new matter state should emerge
at the very beginning of
the shower (not, say, in the middle of it) 
because of the foreseen large mass of the unparticle, 
acting as correlated seeds
of various fragmentation chains.
Notice that a fluctuating number of unparticles 
at the onset of the cascade together with 
a large (and continous) mass spectrum would
induce LRC among the final-state particles.

Let $\rho_1^{(c)}(y_c,y_u)$ 
and $\rho_2^{(c)}(y_{c1},y_{c2},y_{u})$ denote
the one- and two-cluster rapidity densities satisfying
\begin{equation}{\label{eq:rhoc}}
\int dy_c\ \rho_1^{(c)}(y_c,y_u)  =  \langle N_c^u \rangle \ ;\ 
 \int  dy_{c1}dy_{c2}\ \rho_2^{(c)}(y_{c1},y_{c2},y_u) =
\langle N_c^u(N_c^u-1) \rangle
\end{equation} 
where $\langle N_c^u \rangle$ is the average cluster
multiplicity from the initial unparticle, and the variance
is now $D_c^2=\langle (N_c^u)^2 \rangle - \langle N_c^u \rangle^2$.

Defining the single and double differential cross sections for
unparticles in inelastic hadron collisions, 
$(1/\sigma_u)\ d\sigma_u/dy_u$ and $(1/\sigma_u)\ d^2\sigma_u/dy_{u1}dy_{u2}$
respectively, we write for one particle production:
\begin{equation}{\label{eq:decay1u}}
\frac{1}{\sigma_{in}}\frac{d\sigma}{dy}=
\frac{1}{\sigma_u}\ \int dy_udy_c\ \frac{d\sigma_u}{dy_u}\ 
\rho_1^{(c)}(y_c,y_u)\ \rho_1^{(s)}(y,y_{c})
\end{equation}

For two secondaries we have to write this time
\[
\frac{1}{\sigma_{in}}\frac{d^2\sigma}{dy_1dy_2}=
\frac{1}{\sigma_u}\ \int dy_{u1}dy_{u2}dy_{c1}dy_{c2}\    
\frac{d^2\sigma_u}{dy_{u1}dy_{u2}}\ 
\rho_1^{(c)}(y_{c1},y_{u1})\ \rho_1^{(c)}(y_{c2},y_{u2})\ 
\rho_1^{(s)}(y_1,y_{c1})\ \rho_1^{(s)}(y_2,y_{c2})+
\]
\[
\frac{1}{\sigma_u}\int dy_u 
\frac{d\sigma_u}{dy_u}\ \biggr[ \int dy_c\  
\rho_1^{(c)}(y_c,y_u)\ \rho_2^{(s)}(y_1,y_2,y_c)+ \int dy_{c1}dy_{c2}\
\rho_2^{(c)}(y_{c1},y_{c2},y_u)\ \rho_1^{(s)}(y_1,y_{c1})\ 
\rho_1^{(s)}(y_2,y_{c2}) 
\biggr]
\]
where the rhs on the first line corresponds to the emission 
of secondaries from two clusters coming from two
different unparticle sources; on the second line, 
the first (second) term corresponds to
the emission of two particles from a single (two) cluster(s) stemming
from the same unparticle source.
 
Integrating over rapidities of intermediate sources of particles
in the central region, one is led to
\begin{equation}\label{eq:corrunp}
C_2(y_1,y_2)=\langle N_c^u \rangle^2\ (\bar{\rho}_1^{(s)})^2\ D_u^2\ +\ 
\langle N_u \rangle\ (\bar{\rho}_1^{(s)})^2\ D_c^2\ +\ 
\langle N_u \rangle \langle\ N_c^u \rangle\
C_2^{(single)}(y_1,y_2) 
\end{equation}
with $D_u^2=\langle N_u^2 \rangle -\langle N_u \rangle^2$. 
Eq.(\ref{eq:corrunp}) is a generalization of Eq.(\ref{eq:corre}); 
when setting $\langle N_u \rangle =1$ and $D_u^2=0$ 
Eq.(\ref{eq:corre}) is quickly recovered. 

From Eqs. (\ref{eq:shortlongcorr},\ref{eq:corrunp}),   
the short-range part of the 2-particle correlation function 
can be identified now as:
$C_2^{SR}= \langle N_u \rangle \langle\ N_c^u \rangle\
C_2^{(single)}(y_1,y_2)$, 
while the long-range part becomes:
$C_2^{LR}(y_1,y_2)=\langle N_c^u \rangle^2\ (\bar{\rho}_1^{(s)})^2\ D_u^2\ +\ 
\langle N_u \rangle\ (\bar{\rho}_1^{(s)})^2\ D_c^2$. 

Let us note in passing that the 
$\alpha$ and $\beta$ parameters in Eq.(\ref{eq:R2hat})
can be expressed as
\beq\label{eq:alpha2}
\alpha=\frac{1}{\langle N_c \rangle}\ 
\frac{c_s}{(\bar{\rho}_1^{(s)})^2}\ \ ;\ \ 
\beta=\frac{D_c^2}{\langle N_c \rangle^2}
\eeq
in a 2-step scenario, 
while in a 3-step scenario one gets the modified expressions
\beq\label{eq:alpha3}
\alpha=\frac{1}{\langle N_u \rangle \cdot \langle N_c^u \rangle}\ 
\frac{c_s}{(\bar{\rho}_1^{(s)})^2}\ \ ;\ \ 
\beta=\frac{D_u^2}{\langle N_u \rangle^2}\ +\ 
\frac{1}{\langle N_u \rangle}\ \frac{D_c^2}{\langle N_c^u \rangle^2}
\eeq

The expression for $F_2$ is also modified in a three-step scenario
\begin{equation}\label{eq:F2threestep}
F_2=F_2^{(u)}+\frac{F_2^{(c)}}{\langle N_u \rangle}+
\frac{F_2^{(single)}}{\langle N_u \rangle \cdot \langle N_c^u \rangle}
\end{equation}
where we have defined the scaled moment of the
distribution of unparticle sources as
\beq\label{F2u}
F_2^{(u)}=\frac{D_u^2}{\langle N_u \rangle^2}-\frac{1}{\langle N_u \rangle}+1
=\frac{\langle N_u(N_u-1) \rangle}{\langle N_u \rangle^2}
\eeq

\subsubsection{Important remark: Great-Clans}

As argued in \cite{Alexander:2000ux} multi-source structure
in hadron collisions has a clear influence on the correlations among
emitted particles. As the number of independent
sources increases, correlations are diluted, as 
experimentally observed 
in p-A versus p-p and $e^+e^-$ 
collisions \cite{Kittel:2005fu,DeWolf:1995pc,Simic:1979pp}.
Indeed, from Eq.(\ref{eq:F2twostep}) it is apparent that a 
larger $\langle N_c \rangle$, for fixed $F_2^{(single)}$, 
implies a smaller value of $F_2$.
However, notice that 
a quite different situation may happen in a 3-step chain.

Recalling the relation: $F_2=1+1/k$
in a NBD and assuming unparticle/cluster independent
emission ($F_2^{(u)}=F_2^{(c)}=1$),  
it is easy to identify 
\beq\label{eq:k}
K_2=F_2-1=\frac{1}{k} \simeq \frac{F_2^{(single)}}{\langle N_c \rangle}\ \ 
;\ \ \frac{1}{k^{(u)}} \simeq \frac{1}{\langle N_u \rangle}
+\frac{F_2^{(single)}}{\langle N_u \rangle \cdot \langle N_c^u \rangle} 
\eeq
where $1/k(1/k^{(u)})$ is a measure of the aggregation of particles
in clans in a 2-step (3-step) parton avalanche. 
Notice that $k^{(u)}$ turns out to be smaller than $k$ 
(corresponding to a higher aggregation
of decay products according to a clan interpretation 
\cite{Giovannini:2001kj}) provided that 
$\langle N_u \rangle < \langle N_c \rangle$, which
seems a reasonable assumption in our 
cascade picture.

Thus we will refer to a {\em Great-Clan} structure and
consider {\em Great-Clans} \footnote{{\em Super-Clan}, as
an alternative name, may be
misleading since Supersymmetry is
not necessarily concerned here. A {\em Great-Clan}, made of
Clans, recalls metaphorically 
the {\em Great Khan}, Khan of Khans.} 
as, possibly, real physical objects stemming from a HS. 
Correlated particles belonging to a {\em Great-Clan}
would spread over a considerable rapidity interval
(more generally, over a large phase space region) 
which might reach many (say, more than five) 
rapidity units, thus likely involving both $y^*>0$ and
$y^*<0$ hemispheres, of relevance for Forward-Backward
correlation analyses as later commented in more detail.

\begin{figure}[ht!]
\begin{center}
\includegraphics[scale=0.38]{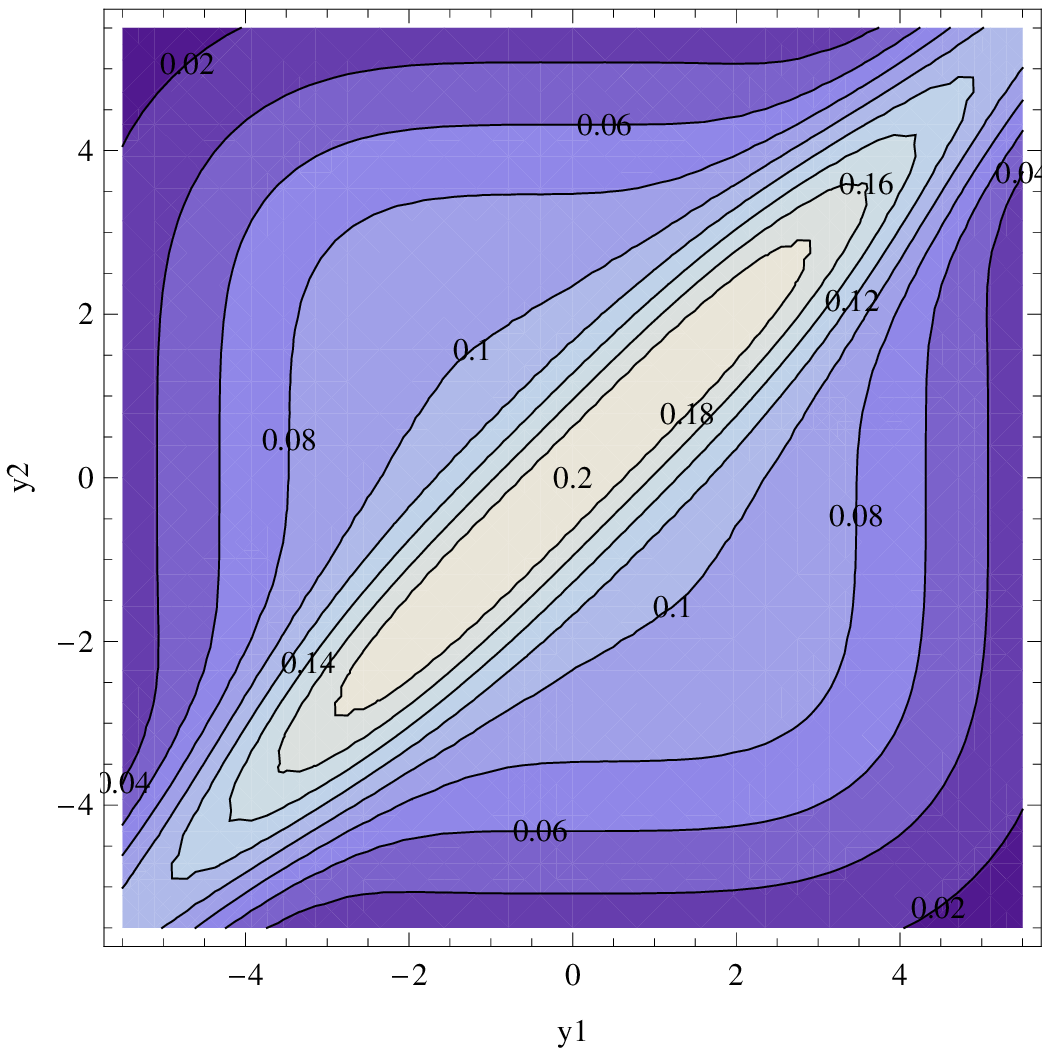}
\includegraphics[scale=0.38]{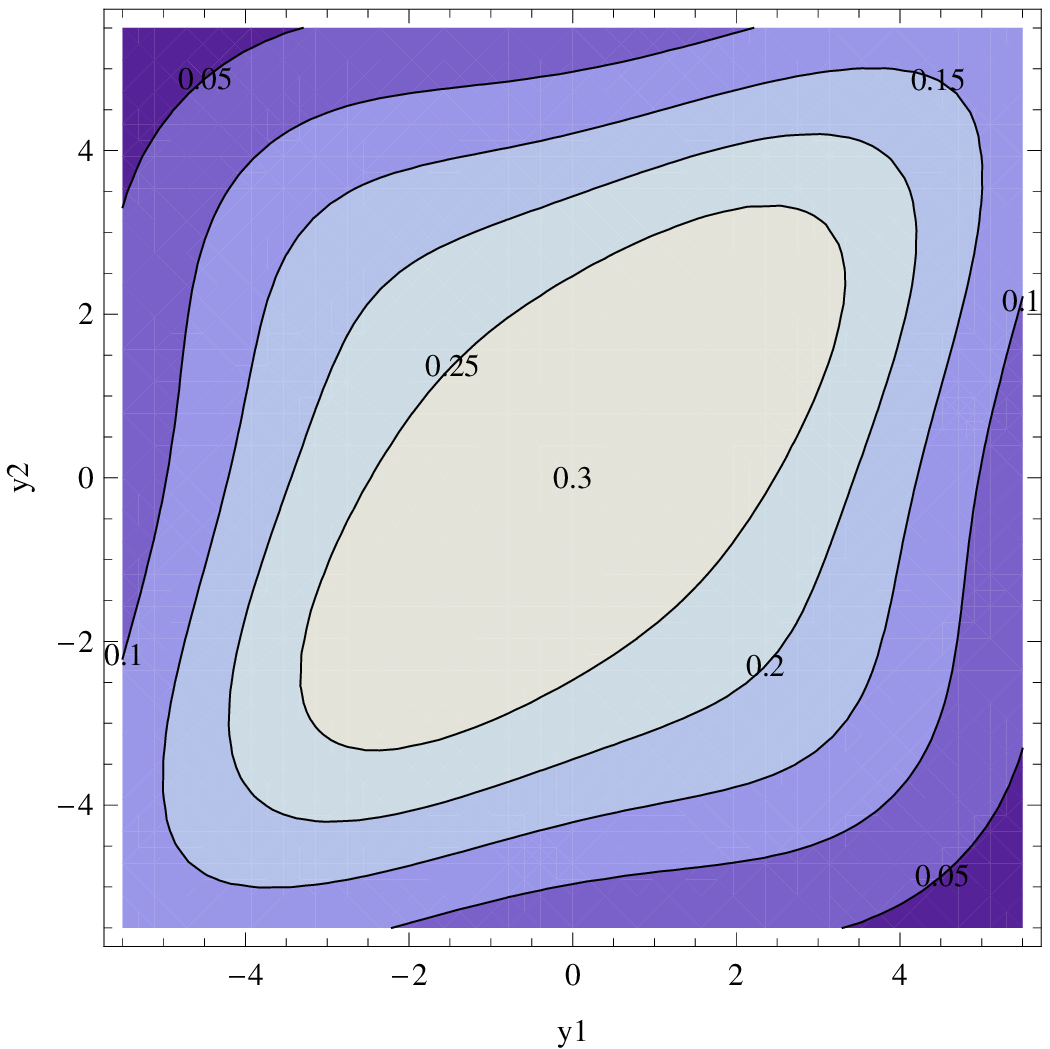}
\includegraphics[scale=0.38]{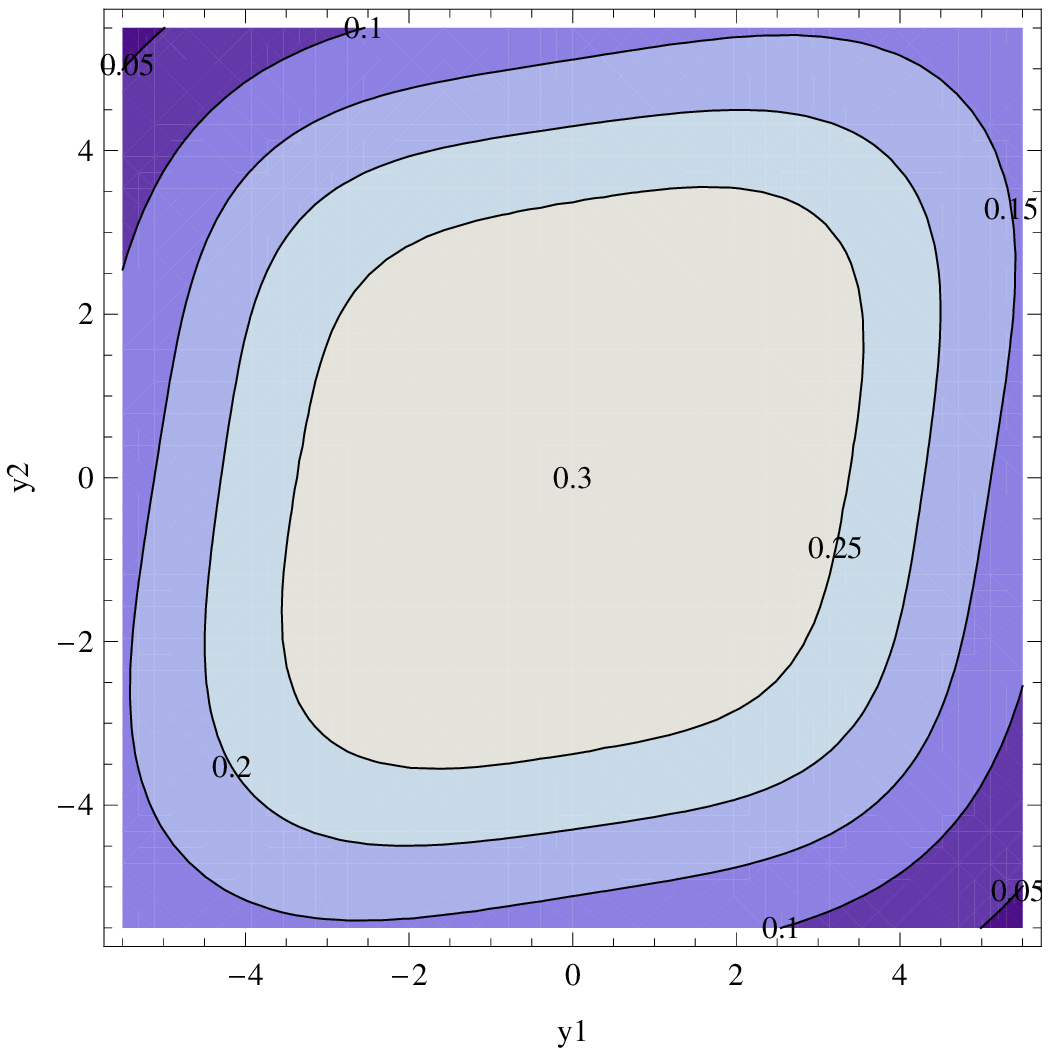}
\includegraphics[scale=0.38]{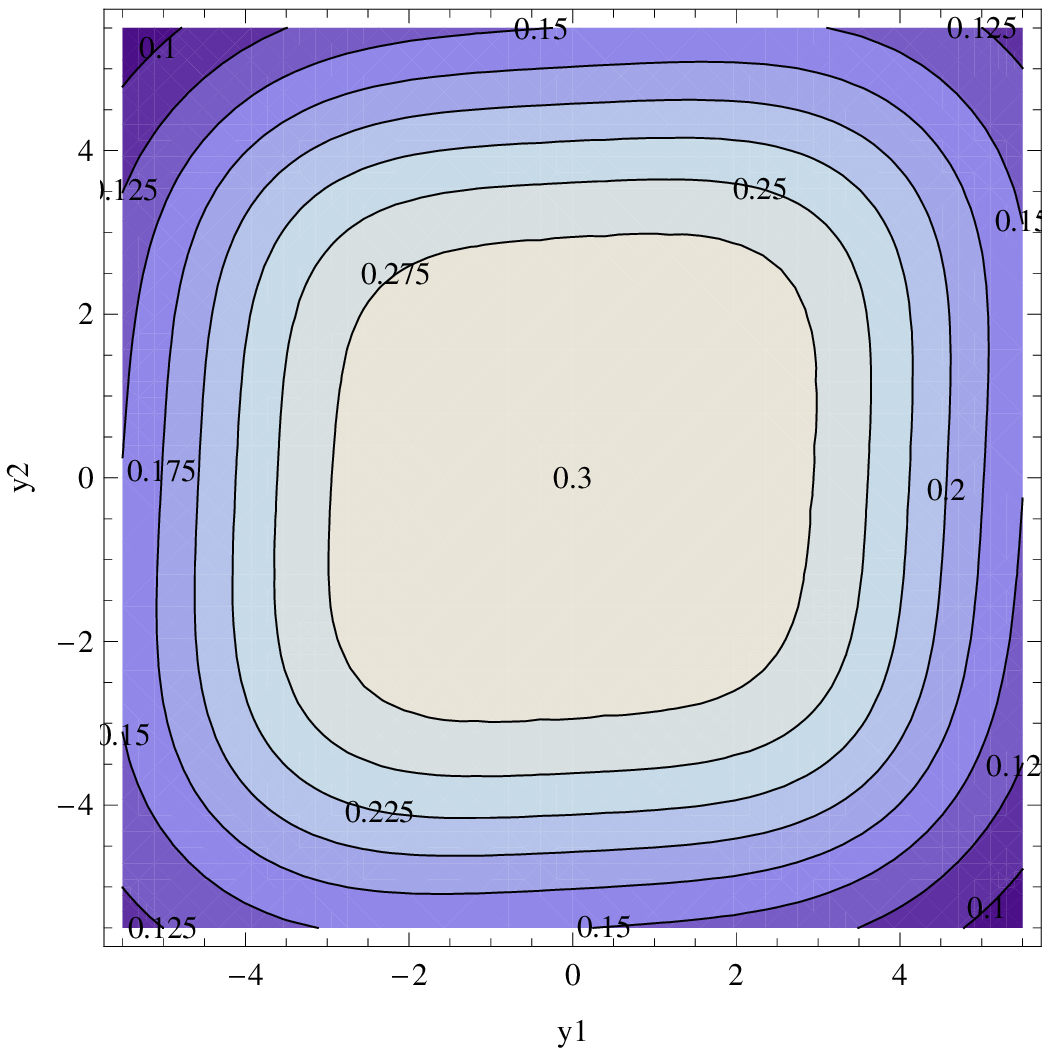}

\caption{Tentative two-dimensional plot of $\hat{R}_2(y_1,y_2)$
in p-p inelastic collision at the LHC energy regime;
from left to right: $\xi_y=0.75,\ 2,\ 5$ and $10$ rapidity
units, respectively. Notice the square-shape of the
contour-lines when $\xi_y$ becomes
comparable to the central plateau length.}
\label{fig:R2}
\end{center}
\end{figure}

\subsection{Inclusive analysis}

Let us remark that the 
enhancement of the LRC found in Eq.(\ref{eq:corrunp}), 
leading to a larger aggregation of particles as
shown in Eq.(\ref{eq:k}), 
comes from terms directly related to a 3-step scenario.
However, this is not the only way as LRC
can be enhanced by a HS.
In fact, it is long known 
that the non-coherent superposition
of two mechanisms in the production cross section (namely inelastic and
diffractive components in p-p collisions) leads
to an enhancement of LRC. This fact, by itself, has no dynamical
content \cite{Foa:1975eu}, although might
be considered as a signal of a 2-component production mechanism.
This point 
is obviously not stressed in the literature because the existence of a
diffractive component in hadronic interactions is 
an already well established fact!
But the situation may of course be very different in 
a discovery strategy of a HS. 

In real p-p data, conventional and (if existing) non-standard 
events would
be mixed in the collected sample; therefore long-distance
correlations caused by a HS 
would naturally arise in inclusive analyses 
of events because
of $\lq\lq$crossed terms'' \cite{Berger:1974vn,Foa:1975eu}
ensuing from the combination of one-particle densities
from each production component.
Indeed, experimental evidence obtained in different 
experiments on p-p and p-A collisions \cite{Foa:1975eu}
suggest that both
$C_2(y_1,y_2)$ and $R_2(y_1,y_2)$ should
exhibit the scaling properties of single-particle distributions
developing a central plateau at asymptotic energies.   

On the grounds of a cluster model and empirical results 
\cite{Foa:1975eu}, $C_2(y_1,y_2)$ can be parametrized as:
\beq\label{eq:C2par}
C_2(y_1,y_2)=Ae^{-|y_1-y_2|^2/4 \delta^2}+B\rho_1(y_1)\rho_1(y_2)
\eeq
where $A$ and $B$ can be roughly estimated from 
our study in the previous section in the central region. 
Similar results follow from the exponential shape in
Eq.(\ref{eq:C2S}).

In Fig.3 we show for illustrative purposes several 2-dimensional plots of 
$\hat{R}_2(y_1,y_2)$ setting different rapidity 
correlation lengths: from $\xi_y=0.75$   
(corresponding to $\delta=0.67$
\footnote{The FWHM of the Gaussian is then about 2 
rapidity units.} in p-p collisions \cite{Alver:2007wy}), 
up to a very large value, $\xi_y =10$ (corresponding
to production of extremely massive
hidden (un)particles). All these
plots are illustrative examples
rather than actual predictions. 

On the other hand, in a conventional multiparticle
production process the 
maximum value of the correlation function
$R_2(0,0)$ is expected to decrease as higher
multiplicities cuts are applied on events,
as proven by experimental results \cite{Breakstone:1982hg}.
An opposite behaviour would hint at NP 
on account of the enhancing LRC term in Eq.(\ref{eq:corrunp})
due to the HS contribution. 
Moreover, let us
note the square-shape of the contour-lines when
the correlation length $\xi_y$ becomes comparable
to the plateau length in the one-particle spectrum,
an early warning about possible NP effects.

\begin{figure}[ht!]
\begin{center}
\includegraphics[scale=0.5]{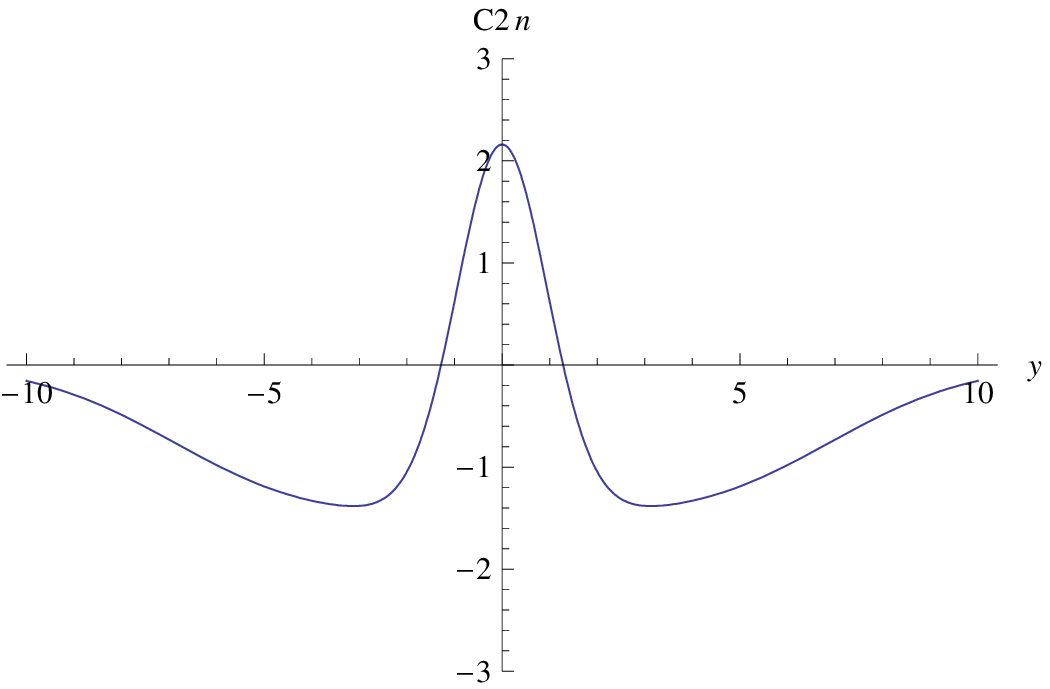}
\includegraphics[scale=0.5]{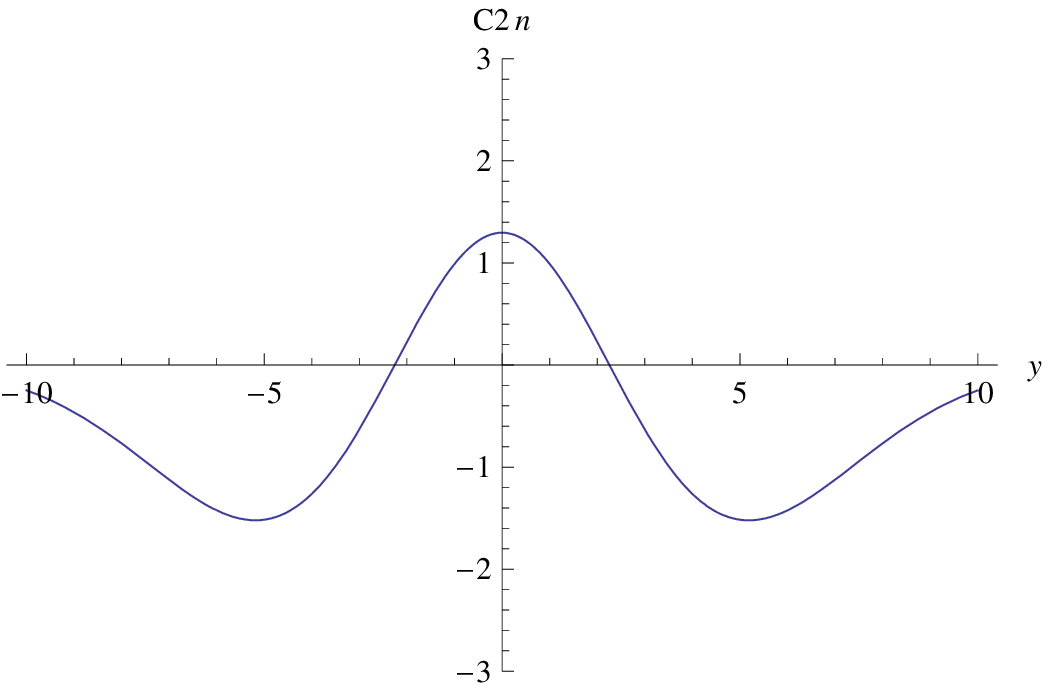}
\includegraphics[scale=0.5]{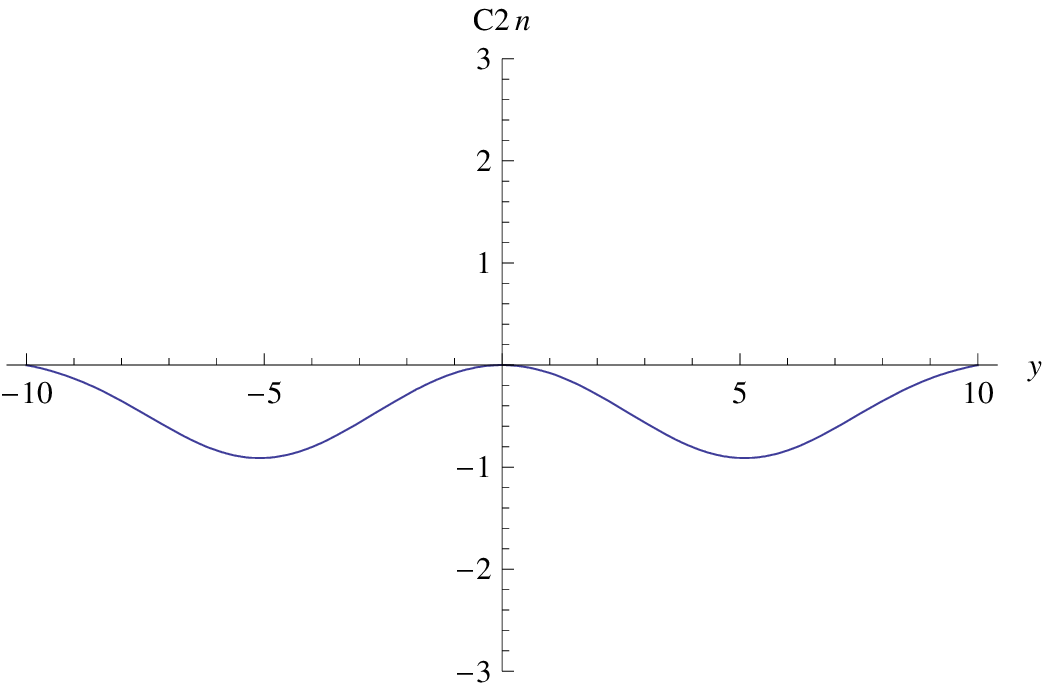}

\caption{{\em Left}: Tentative plot of the 
semi-inclusive $C_2^{(2\langle n \rangle)}(0,y_2)$
function versus $y_2$; 
From left to right: $\delta=0.67$ ($\xi_y=0.75$) ;  
$\delta=1.77$ $(\xi_y=2)$ and $\delta=3.5$ ($\xi_y=4$).}
\label{fig:plotSLL}
\end{center}
\end{figure}

\subsection{Semi-inclusive analysis}

In order to remove as much as possible those non-dynamical
fluctuations caused by unitarity, a {\em semi-inclusive} analysis 
of multiparticle production can be carried out, 
thereby providing an insight into the dynamical 
\footnote{Let us point out, however, that the enhanced 
strength of LRC
due to unitarity ought to be attributed ultimately to 
the underlying dynamics of a HS.} correlations 
of particles emitted by clusters/clans (see, for example, 
\cite{Bell:1983di,Berger:1974vn}) or even larger structures like
{\em Great-Clans} as advocated in this paper.

The semi-inclusive 2-particle correlation
function can be written as \cite{Berger:1974vn,Foa:1975eu}
\begin{equation}\label{eq:C2semi}
C_2^{(n)}(y_1,y_2)= \rho_2^{(n)}(y_1,y_2)-\rho_1^{(n)}(y_1)\rho_1^{(n)}(y_2)
\end{equation}
where we have defined the single $\rho_1^{(n)}(y)$
and double $\rho_2^{(n)}(y_1,y_2)$
rapidity semi-inclusive densities of charged particles
at fixed (charged) multiplicity $n$ in p-p collisions. 

The above definition leads to the following
normalizations:
\beq\label{eq:seminorm}
\int\ dy_2\ C_2^{(n)}(0,y_2)=-\rho_1^{(n)}(0)\ \ ;\ \ 
\int\ dy_1dy_2\ C_2^{(n)}(y_1,y_2)=-n
\eeq
Notice that these conditions force 
$C_2^{(n)}(y_1,y_2)$ to be mostly negative.

In order to estimate the impact
of a foreseen longer correlation length on 
$C_2^{(n)}(y_1,y_2)$, use will be made of
the simple parametrization \cite{Berger:1974vn,Foa:1975eu}:
\beq\label{eq:C2plot} 
C_2^{(n)}(0,y_2)=A_n\ e^{-y_2^2/4\delta^2}-B_n\ \rho_1^{(n)}(0)\ 
\rho_1^{(n)}(y_2)
\eeq

Motivated by Eq.(\ref{eq:C2sl}), we tentatively set $A_n \approx 
c_s\langle N_c \rangle$, while $B_n$ is tuned
in order to satisfy the normalization (\ref{eq:seminorm}).
For definiteness we have 
chosen $n=2 \langle n \rangle$; this choice is motivated by the
reasonable expectation that a HS should rather 
manifest in high-multiplicity events. Notice that 
our actual claim is not providing predictions for
$C_2^{(n)}(y_1,y_2)$, but an educated guess of the 
relative variation of the plotted curve as the correlation
length $\xi_y$ varies. 

In Fig.4 the semi-inclusive correlation
function $C_2^{(2\langle n\rangle)}(0,y_2)$ 
is plotted versus
$y_2$, for different values of the Gaussian parameter
$\delta$ (and $\xi_y$).
A narrow spike around $y_2=0$ generally stands
up over a broad negative pedestal for SRC. 
Notice that longer correlations would 
show up in the plot by smoothing the curve
which may become completely negative along
the rapidity interval. 

A caveat is in order. Assuming a realistic 2-component 
(conventional and non-standard) hadroproduction 
model, two Gaussian/exponential pieces,
with distinct $\xi_{y}$, should be needed 
in both expressions (\ref{eq:C2par}) and (\ref{eq:C2plot})
to fit real data.

\newpage

\section{Forward-Backward correlations in a three-step scenario}

Correlations between the charged-particle multiplicity
in one hemisphere versus that of the other have been
extensively studied in a wide range of center-of-mass energies,
from bubble chamber experiments to LEP and Tevatron
(see \cite{Kittel:2004xr,Kittel:2005fu} and references therein)
for different beams and targets, in particular to understand the dynamics of 
dense matter in heavy-ion collisions 
\cite{Cunqueiro:2006xe,Abreu:2007kv}. 

Following a parallel development, let us introduce
in our analysis the well-known Forward-Backward (F-B)
correlation parameter $b$, defined as 
\begin{equation}\label{eq:b}
\langle n_B \rangle = a + b\ n_F\ \ ;\ \ b\ =\ \frac{D_{FB}^2}{D_{FF}^2}
\end{equation}
where $n_F(n_B)$ represents the multiplicity of the 
forward (backward) hemisphere, respectively, and 
$D^2$ is the corresponding variance, i.e.
\beq\label{eq:fb}
D_{FB}^2 =  \langle n_Fn_B \rangle -\langle n_F \rangle \langle n_B \rangle\
\ ;\ \  
D_{FF}^2 =  \langle n_F^2 \rangle - \langle n_F \rangle^2
\eeq

Assuming no correlation between
the forward and backward hemispheres,
$b$ can be written as \cite{Capella:1978rg} 
\begin{equation}\label{eq:b3}
b=\frac{\int_{y_1>0}\int_{y_2<0} dy_1dy_2\ C_2(y_1,y_2)}
{\int_{y_1>0}\int_{y_2>0} dy_1dy_2\ [C_2(y_1,y_2)+\langle n_F \rangle 
\delta(y_1-y_2)]}
\end{equation}
where the short-range part of $C_2(y_1,y_2)$
should give no contribution in the numerator integral.
Often a gap on the central region is excluded
from the above intervals of integration to focus particularly
on LRC; we will not consider this possibility in our study.

Moreover, a simple expression for $b$ can be obtained in a 2-step scenario,
for the symmetrical situation corresponding to p-p collisions at the LHC, 
when $\langle n_B \rangle = \langle n_F \rangle$; one gets
\begin{equation}\label{eq:b1}
b=\frac{1}{1+\frac{K}{\langle n_F \rangle} \frac{D_s^2}{\langle n_c \rangle}}\ 
\ \Longrightarrow\ 
\ b=\frac{1}{1+\frac{K}{\langle n_F \rangle}}
\end{equation}
where  
$D_s^2=\langle n_c \rangle^2-\langle n_c \rangle^2$ denotes
the variance of the number of particles in single cluster
decays \footnote{Totally equivalent expressions involving
heavy-ion collisions can be found, e.g.,
in \cite{Brogueira:2007ub,Brogueira:2006yk}.}.

The $K$ parameter introduced above is defined through
\begin{equation}\label{eq:K}
\frac{1}{K}=\frac{D_c^2}{\langle N_c \rangle^2}
\end{equation}
where $\langle N_c \rangle$ and $D_c^2$ denote here
the mean number and 
variance of clusters in each hemisphere.

The formula on the left side of Eq.(\ref{eq:b1}) corresponds to  
a Poisson distribution for local particle emission from clusters
(disregarding any hemisphere dependence) 
since then $D_s^2 = \langle n_c \rangle$. Let us note that 
an expression 
similar to (\ref{eq:b1}) is obtained according to the 
Color-Glass-Condensate approach \cite{Armesto:2006bv}
where LRC are in fact expected.

In a 3-step scenario, however, the above formula for $1/K$
is modified becoming
\begin{equation}\label{eq:Kunp}
\frac{1}{K^{(u)}}
=\frac{D_u^2}{\langle N_u \rangle^2}
+\frac{1}{\langle N_u \rangle}\frac{D_c^2}{\langle N_c^u \rangle^2}
\end{equation}
For $\langle N_{u} \rangle =1$ and $D_u^2=0$ 
the expression for $K$ in Eq.(\ref{eq:K}) 
is recovered identifying $\langle N_c \rangle$ and
$\langle N_c^u \rangle$. Let us also note that
Eqs.(\ref{eq:K},\ref{eq:Kunp}) can be put
in correspondence with Eqs. (\ref{eq:alpha2},\ref{eq:alpha3}).

\newpage

Let us stress that, for
Poissonian distributions and not too large  
$\langle N_u \rangle$
(i.e. $ \langle N_u \rangle \lesssim \langle N_c^u \rangle$), 
Eqs.(\ref{eq:K},\ref{eq:Kunp}) imply that
$K^{(u)}$ is {\em smaller} than $K$. We therefore suggest 
that a possible discrepancy found in 
the measurement of $K$, together with other related
observables from F-B correlations,  
in p-p collisions at the LHC 
(see e.g. \cite{Giovannini:2002za} for predictions
within the clan picture) could also shed light on
a hidden (un)particle contribution to the parton cascade. 

F-B correlations have been currently studied for 
different rapidity windows versus the gap size 
separating them. It is striking that in h-h collisions
LRC effects persist up to a gap size of about 6
rapidity units \cite{Kittel:2004xr}. 
Let us remark that very massive states at the onset of
the cascade would imply stronger and even longer distance
correlation effects, leading to larger cluster structures
in rapidity space, thereby motivating
the {\em Great-Clan} concept advocated in this work. 

On the other hand,  as mentioned in section 2.6, 
LRC stemming from Glasma formation
related to the $\lq\lq$ridge'' phenomenon discovered
at RHIC \cite{Dumitru:2008wn} are actually not seen in p-p
collisions, appearing to be unique to nucleus-nucleus
collisions (since in the former case correlations are not
collimated in azimuthal angle). Moreover, a growth of 
the LRC is predicted with the centrality of the collision
in heavy-ion collisions.
In any event, if very long-distance correlations were found in
p-p inelastic collisions at the LHC, a dedicated study
should be certainly carried out, comparing the results 
with RHIC and ALICE data, looking for
distinctive features in order 
to reveal its true origin, either from
a HS, from Glasma formation, or both.

\section{Intermittency and fractal structure of the cascade}

Fluctuations in small phase space regions (intermittency)
have been commonly described by the scaled moments $F_q(\delta y)$, 
where the rapidity interval under study $\Delta Y$
is split into $M$ bins of equal size 
$\delta y=\Delta Y/M$, as already commented in section 4.3. 
Moreover, the fractality nature
of multiparticle hadroproduction (similar to a Cantor dust)
is deeply connected
with intermittent behaviour exhibiting a power-law
dependence of the multiplicity moments with the cell size,
as we shall later see.

For a $\lq\lq$smooth'' (rapidity) distribution,
not showing any fluctuation but the statistical ones,
the moments $F_q(\delta y)$ should become independent of $\delta y$
in the limit $\delta y \to 0$. On the contrary, if
self-similar dynamical fluctuations exist, $F_q(\delta y)$ should
obey a power-law increase at small $\delta y$, i.e.
\begin{equation}{\label{eq:powerlaw}}
\langle F_q(\delta y) \rangle\ \sim\ \delta y^{-\phi_q}\ \sim M^{\phi_q}
\end{equation}
For practical reasons, this power-law scaling (commonly called $M$-scaling) 
is expressed as
\begin{equation}
\ln{F_q}=a_q - \phi_q \ln{\delta y} \equiv A_q+\phi_q \ln{M}
\label{eq:loglaw}
\end{equation}
where $\phi_q$ are called $\lq\lq$intermittency exponents''.
The above power-law dependence of the scaled factorial 
multiplicity moments 
\footnote{Sometimes in the literature, intermittency may refer
to any increase of the factorial moments with decreasing phase space intervals
without need of a scaling behaviour as shown in Eq.(\ref{eq:powerlaw}). 
The extended version just
indicates positive correlations in rapidity
which increase at smaller bins. However, let us 
stress the relevance of the original proposal of intermittency 
connected with fractality through 
statistical self-similarity along the full
cascade process.} 
on the bin size $\delta y$ is, in fact, 
a signature of self-similarity in the fluctuation
pattern of the particle multiplicity \cite{Dremin:1989wh}.

In reality, particle correlations among emitted particles
and intermittency could (should) also be studied 
using the normalized factorial cumulants $K_q$ (see footnote $\#8$)
instead of $F_q$, as for example carried out by OPAL
in the study of 
hadronic $Z$ decays \cite{Abbiendi:2006qr,Abbiendi:2001bu}.
Once experimental data 
become available from LHC experiments, the analysis of factorial cumulants
will become compulsory in p-p inelastic events. Meanwhile, we
restrict our study to the power-law behaviour
given by (\ref{eq:powerlaw}).

\begin{figure}[ht!]
\begin{center}
\includegraphics[scale=0.5]{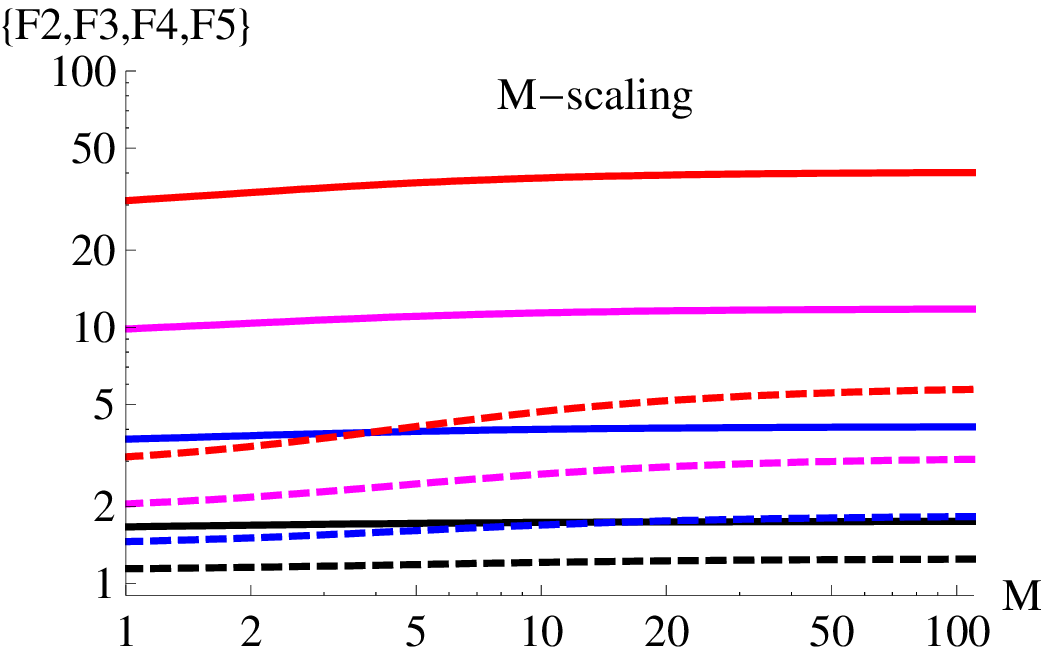}
\includegraphics[scale=0.5]{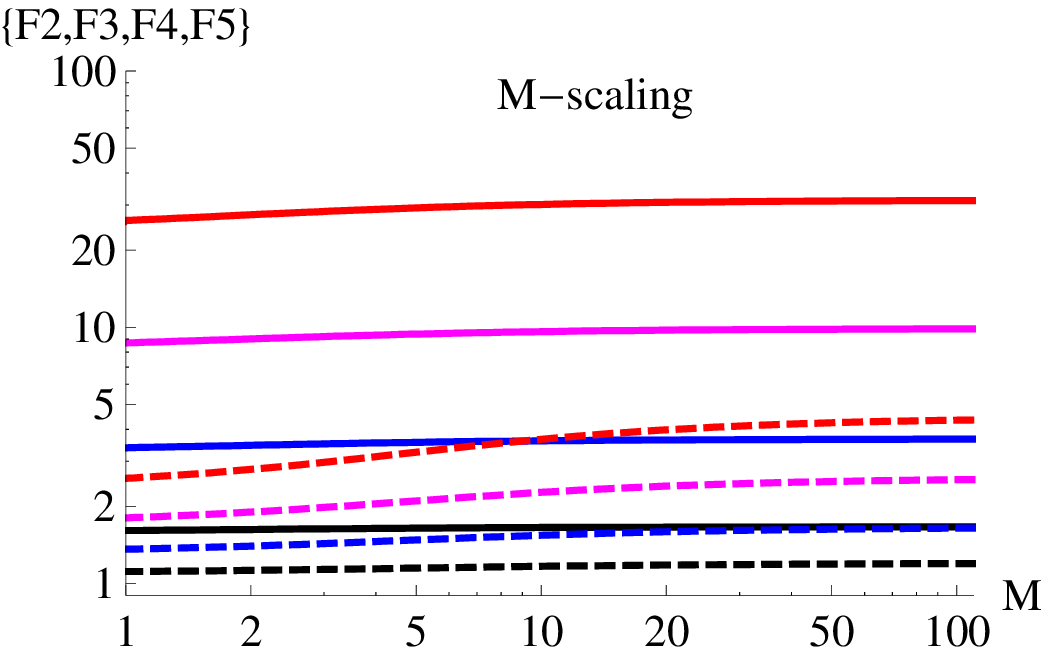}
\includegraphics[scale=0.5]{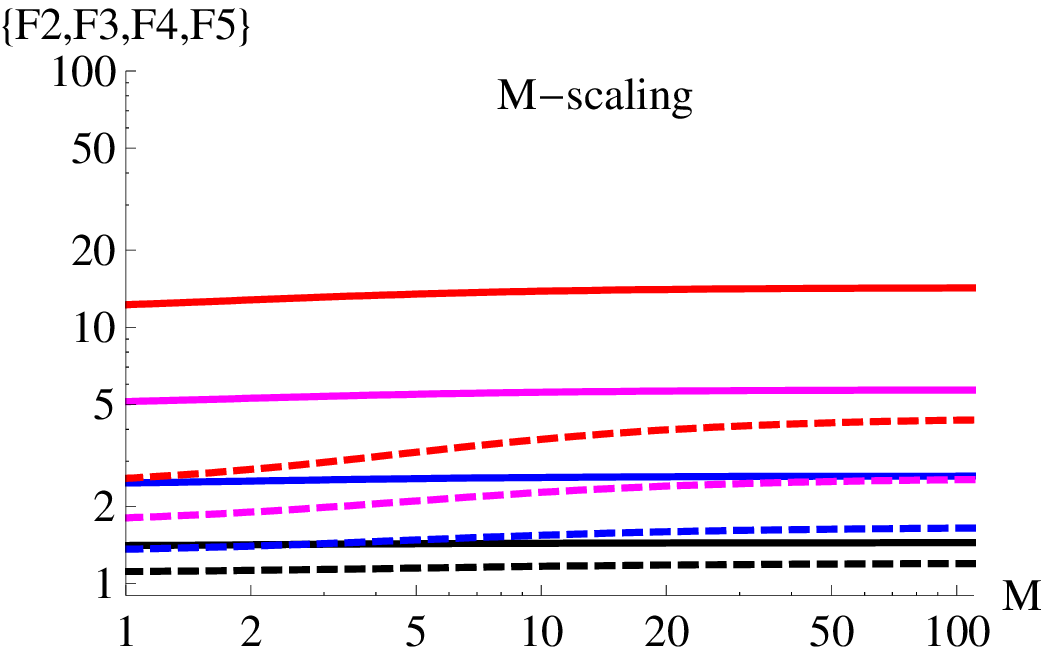}
\caption{Log-log plots of the factorial normalized moments.
From down upwards:
$F_2$ (black), $F_3$ (blue), $F_4$ (magenta), and
$F_5$ (red) curves, versus $M$. Dashed lines
correspond to a 2-step scenario obtained
from Eq.(\ref{eq:F2twostep}), while solid lines correspond to
a 3-step scenario obtained from Eq.(\ref{eq:F2threestep}). 
{\em a) Left panel}: $\langle N_c \rangle=8$, and $\langle N_u \rangle=2$,  
$\langle N_c^u \rangle =4$, assuming all
distributions to be Poisson: $F_2^{(u)}=F_2^{(c)}=1$;
{\em b) Middle panel}: The same setting $\langle N_c \rangle=10$, and
$\langle N_u \rangle =2$, $\langle N_c^u \rangle =6$;
{\em c) Right panel}: The same setting $\langle N_c \rangle =10$, and
$\langle N_u \rangle=3$, $\langle N_c^u \rangle = 6$}
\label{fig:F2}
\end{center}
\end{figure}

\subsection{Numerical estimates of normalized factorial moments}

Our main concern 
in this section is to investigate about 
the multi- or mono-fractal nature of the 
showering process in a two-step scenario
versus three-step scenario. We shall study 
the intermittency foreseen for
multiplicity distributions in p-p collisions. 
Our goal is by no means
giving $\lq\lq$predictions'', rather we want
to assess the sensitivity of some {\em fractality
estimators} to a hypothetical non-standard step in the 
hadroproduction cascade, once experimental data from
LHC become available. To this aim,  
we will rely upon
the estimates of $F_2$ from Eqs.(\ref{eq:F2twostep}) and 
(\ref{eq:F2threestep}), and on    
the scaling relations of Eq.(\ref{eq:Fqrelations})
for the $F_q$  moments of higher rank.

Let us first compute $F_2^{(single)}$ by making the assumption
that $C_2^{(single)}$  and thereby  $C_2^{SR}$ display
the exponential form of Eqs.(\ref{eq:C2single},\ref{eq:C2S}). 
Let us remark that whenever
we consider a 3-step scenario, we extend
the concept of cluster up 
to a {\em Great-Clan}, i.e. quite more final-state particles 
are correlated than in a single cluster
assuming that the effective correlation length becomes larger. 

Upon integration of Eq.(\ref{eq:F2single}) one gets
\begin{equation}
F_2^{(single)}(M)=1+\frac{c_s}{(\bar{\rho_1}^{(s)})^2}\ G_{\xi_y}(M/m) 
\label{eq:F2corr}
\end{equation}
where $m=\Delta Y/\xi_y$, and
\begin{equation}\label{eq:g}
G_{\xi_y}(r)=2\ [r-r^2+r^2e^{-1/r}]
\end{equation} 
Note that $G_{\xi_y} \to 2\xi_y/\delta y$ for
small $M$ (i.e. $\delta y >> \xi_y$), and $G_{\xi_y} \to 1$ 
for large $M$ (i.e. $\delta y \to 0$).

In our toy model, the LRC term was assumed to be a constant, so
the resulting (multi)fractality 
derives from the single cluster decay via
the behaviour of $G_{\xi_y}(r)$ as a
function of $M$ or $\delta y$. The limit when $F_2^{(single)}$
is negligible as compared to such a constant LRC term 
implies in fact nonfractality.

In Fig.5 we show a set of double logarithmic 
plots for several normalized factorial moments of the
multiplicity distribution,
$F_2,F_3,F_4$ and $F_5$ (from down to up), versus the 
(pseudo)rapidity binning $M$.
The curves have been naively calculated by assuming
Poisson-like distributions for all particle sources.
Dashed (solid) lines refer to expectations from a conventional
two-step (three-step) cascade
employing Eqs.(\ref{eq:F2twostep}) and (\ref{eq:F2threestep})
with different assignments for the
average numbers of unparticles and clusters. Besides we assumed
$\xi_y=0.75$ and $\xi_y=2$ for the 2- and 3-step cascades,
respectively.

\newpage

The examples shown in Fig.5 correspond to:
\begin{itemize}
\item[\bf a)]  $\langle N_c \rangle=8$, and $\langle N_u \rangle=2$, 
$\langle N_c^u \rangle=4$
\item[\bf b)] $\langle N_c \rangle=10$, and $\langle N_u \rangle =2,\ 
\langle N_c^u \rangle=6$
\item[\bf c)] $\langle N_c\rangle=10$, and 
$\langle N_u \rangle=3,\ \langle N_c^u \rangle=6$.
\end{itemize}

In reality, the bin size dependence of $F_q$ does not follow
a power-like law in the whole rapidity interval, but
can be fitted separately by power laws for large/small bins. 
Hereafter we restrict
our attention to the more or less linear region
between $M=1$ and $M=20$ before the levelling-off.

From inspection of these plots we can conclude that:
\begin{itemize} 

\item The height of all curves may vary significantly 
when passing from one scenario to another.
Although an increase or decrease of the $F_q$ asymptotic values 
(i.e. at large $M$) may result depending on the 
assumptions for $\langle N_u \rangle$, $\langle N_c^u \rangle$ 
and $\langle N_c \rangle$, higher curves generally stem from
for the 3-step chain, provided that $\langle N_u \rangle \lesssim
\langle N_c\rangle$, as already
pointed out in section 5.2.1.

\item The slopes  $\phi_q$ differ appreciably when
comparing the two-step against the tree-step cascade, especially for
higher $q$ values. This fact could be anticipated because of
the expected larger $C_2^{LR}$ constant contribution to the 
correlation function in the latter case. 
As we shall see, the multifractality revealed by
the $q$-dependence of $d_q$ weakens accordingly.

\end{itemize}

The intermittent behaviour observed by many experiments
(see e.g.\cite{Kittel:2005fu,DeWolf:1995pc,Capella:1989xg,
Abbiendi:2001bu,Albajar:1992hr,Sarkisian:1994xa,Ahmad:2006sb,Li:2007zzt}) 
in the analysis of factorial moments of spectra of emitted
secondaries indicate the existence of some multifractal structure
of the production process. We will address this issue
in the following section, using the $\lq\lq$results'' 
from our toy model shown in Fig.5.

\subsection{Fractality}

As mentioned above, hadronization exhibits a self-similarity pattern and
therefore a multifractal structure: the pictorial description
of jets within jets, within jets is commonly employed
to illustrate this behaviour. In this regard, 
the Lund group \cite{Dahlqvist:1989yc}
presented a suggestive interpretation of the
cascade based on a colour dipole picture by means of a 
multi-faceted surface due to the consecutive emission
of many gluons.
Because of its iterative nature, the process generates
a Koch-type fractal curve at the base-line, whose length
is proportional to the particle multiplicity. This
curve becomes longer when analized with higher resolution:
it is indeed a fractal curve characterized by the fractal dimension
$d_f = 1 + \gamma_0$, where $\gamma_0=\sqrt{3\alpha_s/2\pi}$ stands
for the QCD anomalous dimension.

Accordingly, the multiplicity of emitted particles
in small phase space intervals 
should show intermittent features, 
with multifractal (R\'enyi) dimension 
$D_q \simeq 2 \gamma_0$ for large $q$. 
The multifractal dimension $D_q$ and 
the anomalous fractal dimension  
$d_q$ are related through:
$D_q=1-d_q$  \cite{Ochs:1992tg}; in turn $d_q$, which 
is a measure for the deviation of $D_q$ 
from an integer, can be related to the intermittency coefficient
$\phi_q$ as:\cite{Lipa:1989yh,Hwa:1989vn}
\beq\label{eq:dq}
d_q=\frac{\phi_q}{q-1}
\eeq

It has been widely argued in the literature 
since Ref.\cite{Bialas:1990xd}
that if QGP is created in hadronic collisions,
a phase transition to hadron matter must take place.
If such a transition is close to second-order 
the outgoing hadron system should show 
intermittency with $d_q$
weakly depending on $q$ in contrast to 
an approximate linear dependence when a cascade 
process occurs.

\begin{figure}[ht!]
\begin{center}
\includegraphics[scale=0.5]{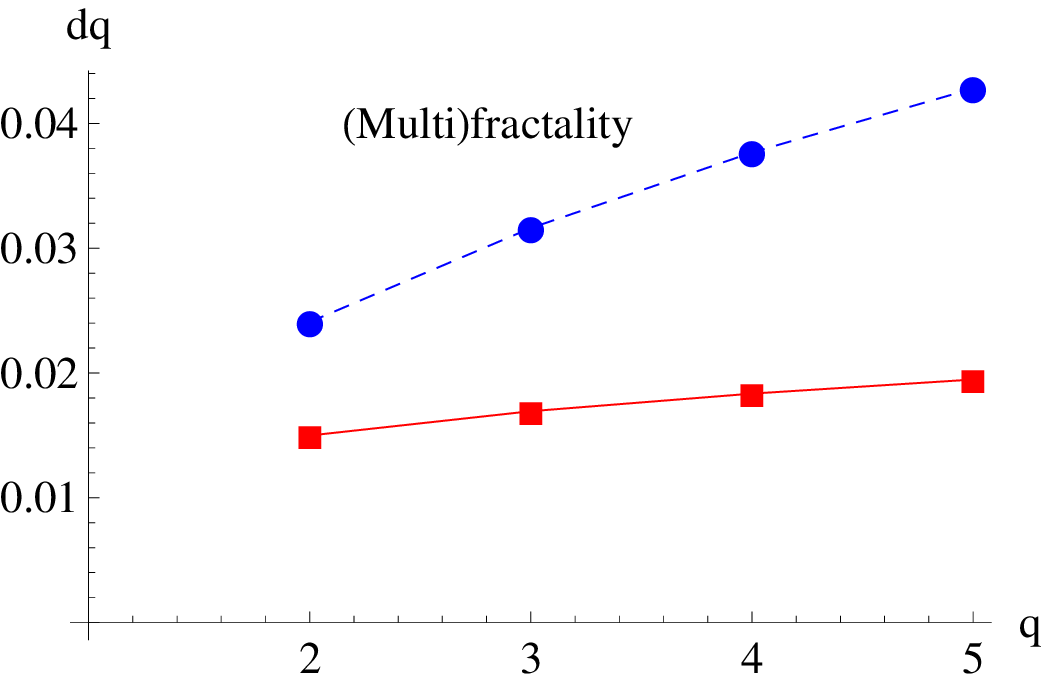}
\includegraphics[scale=0.5]{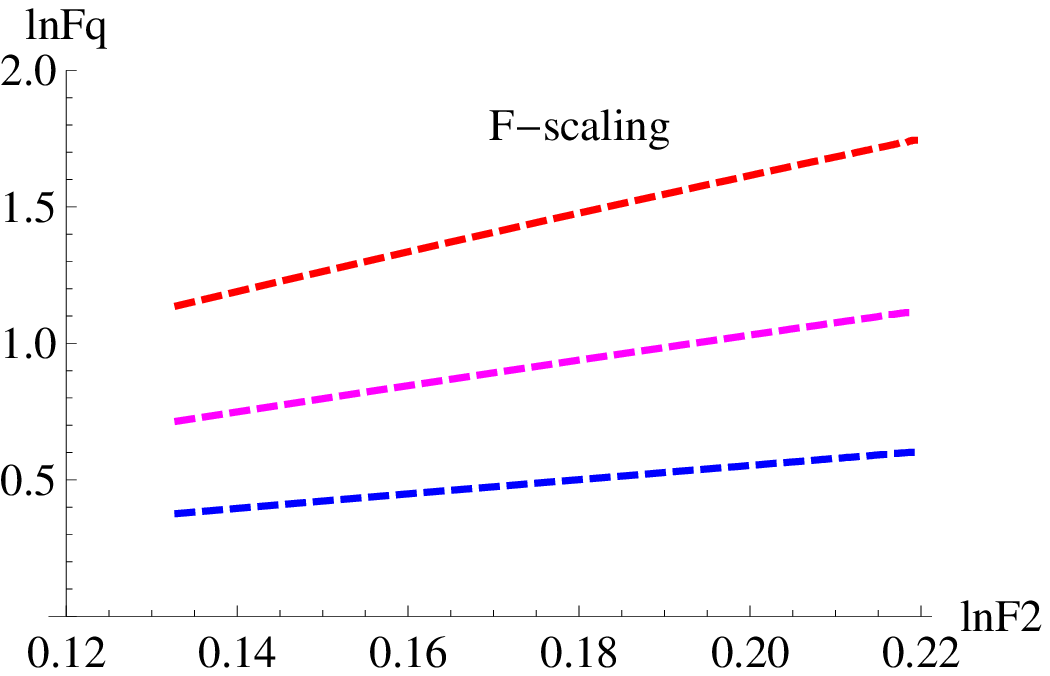}
\includegraphics[scale=0.5]{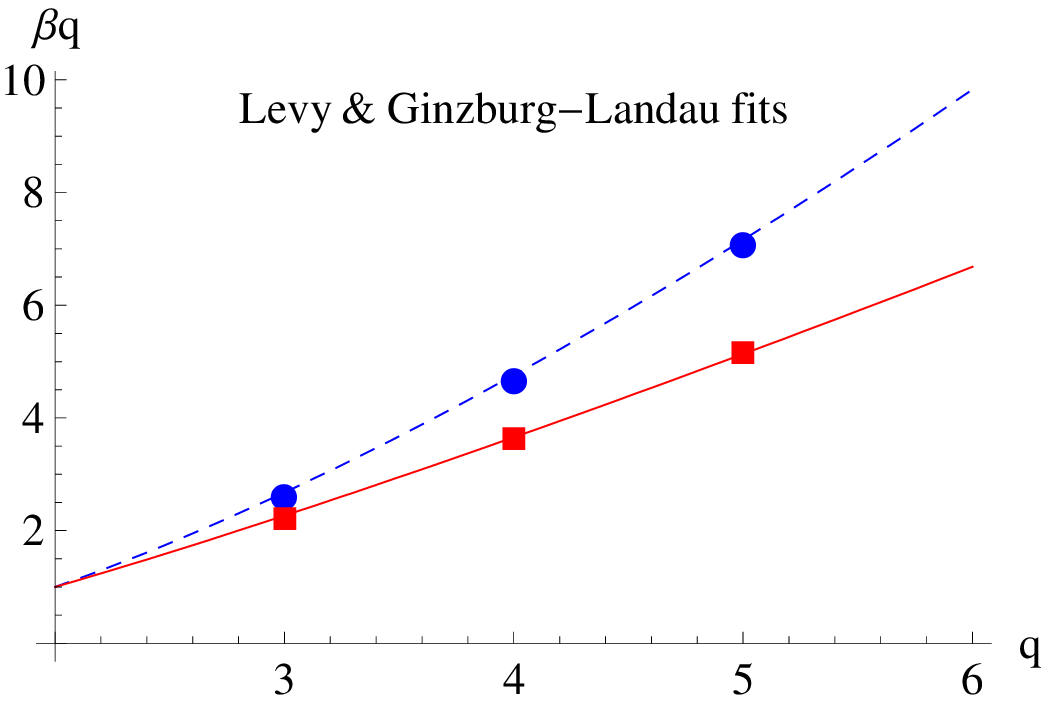}
\caption{Study of the fractality character of the parton cascade.
{\em Left}: $d_q$ versus $q$ where the dashed blue line
stands for a 2-step scenario with $\langle N_c \rangle=8$, while
the solid red line stands for a 2-step scenario with  
$\langle N_u \rangle =2$, $\langle N_c^u \rangle=4$; 
{\em Middle}: $\ln{F_q}$ versus
$\ln{F_2}$  for $q=3,4,5$ (from down upwards) 
; {\em Right:} $\beta_q$ versus $q$ and fits using either
Eq.(\ref{eq:Levy}) or Eq.(\ref{eq:nu}) in
a conventional cascade 
with $\mu=1.46$, $\nu=1.42$, or a three-step cascade with
$\mu=0.72$, $\nu=1.18$, respectively} 
\label{fig:Dq}
\end{center}
\end{figure}

Let us remark that for
a second-order phase transition from plasma to hadrons 
the correlation length should be larger than the size
of the largest system produced.
This indeed resembles the claim of this paper 
when a new state of matter is
assumed at the onset of the showering.

In Fig.6 we show the 
behaviour of $d_q$ for $q=2,3,4,5$ obtained from our 
toy analysis based on the (pseudo)rapidity binning.
One can quickly see that the $q$-dependence of $d_q$ weakens
in the three-step scenario. This is in fact a general feature
for almost every reasonable parameter combination in our study
as could be expected as a consequence of long-range correlations
(both from $C_2^{LR}$, and $C_S^{SR}$ associated to  
a larger $\xi_y$) yielding stronger contributions
to the $F_q$ moments.  
This conclusion agrees with the experimental
finding that the anomalous dimension generally decreases
for a growing complexity of the process (e.g. from
muon-proton, p-p, to p-A and A-A interactions), theoretically
accounted for in \cite{Lipa:1989yh}.

\subsection{Ratios of intermittency exponents}

Another interesting feature closely related to the
above discussion is the power scaling between 
$F_q$ and $F_2$, i.e. 
\beq\label{eq:Fscaling}
F_q(M) \sim [F_2(M)]^{\beta_q}
\eeq
commonly known as $F$-scaling, obviously implying
\begin{equation}\label{Fscaling}
\ln{F_q}(M)=\beta_q\ln{F_2}(M)
\eeq
where the $\beta_q$ coefficient can be expressed as
\beq\label{eq:beta} 
\beta_q=\frac{\phi_q}{\phi_2}=\frac{d_q}{d_2}\dot (q-1)
\eeq
Notice that the validity of $M$-scaling as
shown in Eq.(\ref{eq:powerlaw})
guarantees the validity of $F$-scaling, but  
Eq.(\ref{eq:Fscaling})
may be valid even if Eq.(\ref{eq:powerlaw}) is not. 

In Fig.6 (middle), the corresponding lines derived from our toy model
are drawn for $q=3,4,5$. Monofractality ($d_q=d_2,\ \forall q$)
of course implies $\beta_q=q-1$.

As mentioned previously, one could (should) consider the
factorial cumulant $K_q$ as a function of $K_2$
in a more realistic study, or when using experimental data. 
In our case, no additional
information would be really added following up from such 
study due to the simplicity of our framework.

\newpage

\begin{table*}[hbt]
\setlength{\tabcolsep}{0.4pc}
\caption{Values of $\beta_q$, $\mu$ and $\nu$ 
obtained in our toy model for the 
two- and three-step scenarios using $\langle N_c \rangle=8$,
and $\langle N_u \rangle=2$, $\langle N_c^u \rangle=4$, respectively.
The result of the fit employing 
the L\'evy law given in Eq.(\ref{eq:Levy}) is
represented in Fig.6 (right) for both scenarios, 
yielding the values of $\mu$ shown below.}

\label{FACTORES}

\begin{center}
\begin{tabular}{cccccc}
\hline
Scenario & $\beta_3$ & $\beta_4$ & $\beta_5$ &  $\mu$ & $\nu$ \\
\hline
2-step & $2.63$ & $4.69$ & $7.11$ & $1.46$ & $1.42$ \\
\hline
3-step & $2.26$ & $3.67$ & $5.20$ & $0.72$ & $1.18$ \\
\hline
\end{tabular}
\end{center}
\end{table*}

\subsubsection{L\'evy stable law description}

Brax and Peschanski \cite{Brax:1990jv} proposed
a better approximation than  Eq.(\ref{eq:beta})
using a L\'evy stable law description of multiparticle
production. Hence instead of Eq(\ref{eq:beta})
the $q$-dependence of $\beta_q$ is given by: 
\beq\label{eq:Levy}
\beta_q=\frac{\phi_q}{\phi_2}=\frac{q^{\mu}-q}{2^{\mu}-2}
\eeq
where $\mu$ is the L\'evy index (also known as the
degree of multifractality) which allows an estimation
of the cascading rate; $\mu$ should be
in principle restricted to the interval $0 < \mu \leq 2$ 
(region of stability). Some violations of this limit have been found
in the analysis of experimental data in h-h collisions 
\cite{Kittel:2004xr,Kittel:2005fu}, though can be adequately 
reinterpreted by invoking high anisotropy of phase space
in multiparticle production \cite{Zhang:1995uf}. 

The introduction of the parameter $\mu$ helps 
classifying intermittency patterns due to different kinds 
of phase transitions, either thermal $\mu<1$ (e.g. quark-gluon plasma), 
or non-thermal $\mu>1$ (cascading process).

Two extreme cases can be distinguished:
\begin{itemize}
\item $\mu=2$: yielding $\beta_q=q(q-1)/2$, i.e. $d_q/d_2=q/2$, 
corresponding to a self-similar branching process.

\item $\mu=0$: yielding $\beta_q=q-1$, i.e. $d_q/d_2=1$, the
system displays a monofractal behaviour. 
\end{itemize}

Whereas cascade models show a multifractal structure,
a hadronic system undergoing a second-order transition
should lead to intermittency associated to a monofractal
structure. It is well-known that the
experimental measurement of this coefficient
in relativistic heavy-ion collisions should be useful in
detecting the formation of QGP \cite{Tawfik:2001az}.

Motivated by the above remarks, we have studied using our toy model 
the $F$-scaling in p-p inelastic
collisions under the hypothesis of an extra stage from a HS.
In Table 1 we present the values obtained for $\beta_q$ 
corresponding to both scenarios, 
providing two quite different values for
the $\mu$ parameter: $\mu=1.46$ obtained for a 
conventional fragmentation chain, versus
$\mu=0.72$ once a new state of matter has been
added into the cascade. Our results are consistent
with the interpretation that the
production process approaches monofractality
(or merely a loss of fractality as all $d_q$ values become
smaller as can be seen from Fig.6) when a HS
modifies the onset of the parton cascade.
Note that unparticle stuff in particular could display
itself a fractal nature acting as a fluctuating mass
source \footnote{A fractal structure of the source in space-time 
is discussed in \cite{Bialas:1992ca,Utyuzh:1999eb}}
of (clusters of) particles 
in the hadronization chain, 
on account of its continous mass spectrum.

Let us remark however that a full exploration of the
fractal properties of multiparticle production actually requires
a 3-dimensional analysis once experimental data become available
at the LHC.

\newpage

\begin{table*}[hbt]
\setlength{\tabcolsep}{0.4pc}
\caption{$\mu$ and $\nu$ values 
obtained in our toy model using different
combinations of $\langle N_u \rangle$, $\langle N_c^u \rangle$ 
in a three-step cascade (right side), versus a two-step cascade 
(left side) for different $\langle N_c \rangle$ values.}

\label{FACTORES2}

\begin{center}
\begin{tabular}{|c|c|c||c|c|c|c|c|}
\hline
$\langle N_c \rangle$ &  $\mu$ & $\nu$ & $\langle N_u \rangle$  &
$\langle N_c^u \rangle$ &  $\langle N_u \rangle \cdot \langle N_c^u \rangle$
& $\mu$ & $\nu$ \\
\hline
2 & 0.72  & 1.18 &  2  & 2 & 4 & 0.60 & 1.15  \\
\hline
4 & 1.10  & 1.30 &  1  & 4 & 4 & 0.43 & 1.10  \\
\hline
6 & 1.29  & 1.37 &  3 &  4 & 12 &  0.98  & 1.26 \\
\hline
8 & 1.46  & 1.42 &  2 &  4 & 8 &  0.72  & 1.18 \\
\hline
10 & 1.50  & 1.45 &  1 &  10 & 10 &  0.50  & 1.12 \\
\hline
18 & 1.69  & 1.52 &  3 &  6 & 18 & 1.01 & 1.27 \\
\hline
24 & 1.75  & 1.55 &  2 &  10 & 20 &  0.85  & 1.22 \\
\hline
\end{tabular}
\end{center}
\end{table*}

\subsubsection{Ginzburg-Landau description}

The Ginzburg-Landau (GL) theory \cite{Ginzburg:1950sr}
has been applied to a wide
variety of phase state transition phenomena, from
superconductivity to quark-gluon matter 
\cite{Abbott:1995as}.  
Here we seek a (hopefully fruitful) analogy between our approach 
to the quest for a HS
in p-p inelastic collisions, and a GL description
of the deconfining phase transition from QGP to hadronic matter
\cite{Hwa:1992uq,Cao:1996tg}. 

Instead of the law expressed in Eq.(\ref{eq:Levy}), $\beta_q$
is parametrized now as 
\beq\label{eq:nu}
\beta_q=(q-1)^{\nu}
\eeq 
where $\nu=1$ characterizes the critical phase transition,
ordinarily implying monofractality. The $\lq\lq$universal'' 
exponent was computed analytically in \cite{Hwa:1992uq} 
for a second order transition, finding $\nu=1.3$.

Once again, 
we perform a comparison between the expected 
$\nu$ value according to a standard cascade versus
the value obtained under the assumption of 
an extra stage associated to a HS, based on our toy model. 
In Fig.6 (right) we show the result of the 
power-fit Eq.(\ref{eq:nu})
of $\beta_q$ as a function of $q$
obtained from $\lq\lq$data'' collected in Table for 
in both scenarios.
In this particular example, we find $\nu=1.42$   
and $\nu=1.18$, for the two- and three-step cascade, 
respectively.

In addition, Table 2 shows several more couples of $\mu$ and $\nu$ 
exponents, to be used as quantitative estimators
of the fractal character of the cascade
obtained for different parameter combinations.
Let us stress again that our goal is not making
predicitions on $\mu$ and $\nu$, but estimating the relative
variation in both scenarios.
One can indeed conclude that both $\mu$ and $\nu$ values turn out to be
systematically smaller in a 3-step cascade
than in a 2-step scenario, remarkably even when
$\langle N_c \rangle \leq \langle N_u \rangle \cdot \langle N_c^u \rangle$. 
Notice, however, the increase of $\mu$ and $\nu$ with 
a growing $\langle N_u \rangle$, i.e. the 
average number of (un)particles from a HS involved in the cascade.
Hopefully, the foreseen differences 
will allow to distinguish experimentally
between the two scenarios
using p-p data to be collected soon at the LHC \cite{lhc}.

\subsubsection{Discussion}

Since multifractality is generally viewed as 
a manifestation of a self-similar random cascade 
(i.e. a non-thermal process), it may seem 
striking that adding a new
step to the parton avalanche would imply the
opposite effect, i.e. getting closer to
monofractality usually associated with (second-order) 
thermal transitions in heavy-ion collisions. 
This apparently paradoxical result could be seen, however, 
in accordance with the existence of a larger 
correlation length in a three-step scenario with heavy hadronic 
objects initiating the hadronization chain. The
new state of matter would, in an effective way, play 
an analagous role to that expected 
for QGP in central heavy-ion collisions.

\section{Summary and prospects}

In hadron-induced reactions, multiparticle production
is generally described as proceeding through a two-step
process. First the interaction between the colliding 
constituents (active partons) gives rise to
strings/clusters/clans/fireballs, subsequently 
decaying through a QCD cascade and  
soft hadronization into final-state SM particles. Of course,
hadroproduction is actually a much more complex process
because of extra event activity caused by 
parton remnants, initial and final-state radiation, etc.

New physics associated to higher mass scales is 
generally expected to show up
in high-$p_{\bot}$ events. In this work, however, 
we have focused on rather diffuse soft
signals in p-p inelastic interactions, 
likely tagged by hard decay products. 
We argued that
a non-standard state of matter from a HS (e.g. unparticle stuff)
would give rise to a three-step cascade in hadron collisions, 
altering observables related to rapidity particle correlations  
which can be determined at the LHC to a large accuracy.

On the one hand, two-particle (or higher) 
rapidity correlation
functions are sensitive to a longer correlation
length caused by either fluctuations in the number of
particle sources, or (very) massive
intermediate state decays.
Expectedly both $C_2(y_1,y_2)$ and $R_2(y_1,y_2)$
functions should develop a plateau in a two-dimensional plot
along the diagonal $y_1=y_2$
of similar length as the single-particle spectrum,
furnishing a $rule$ (or scale) 
to determine dynamical correlations, e.g., along the 
perpendicular diagonal 
$y_1=-y_2$. Naively, a square-shape of contour-lines 
at mid-rapidities and an increase
of $R_2(0,0)$ with event multiplicity might
hint at NP.
Moreover, since CGC/Glasma also predicts LRC, 
a comparative study between A-A collisions 
and p-p collisions should be performed to find out 
distinctive signatures of a HS
from extended rapidity sources expected in QCD.

By fixing the multiplicity of events, those LRC due to unitarity constraints 
should be largely suppressed as the different sources
(at any stage) are not yet allowed to fluctuate freely.
We have checked that a semi-inclusive analysis could
therefore be useful to determine a correlation length likely
longer than expected in a conventional parton showering,
due to the large size of clans (dubbed {\em Great-Clans})
stemming from a HS. 
We also performed a study of forward-backward correlations
concluding again that a HS could modify the correlation parameter
usually defined for such analyses.

On the other hand, the afore-mentioned three-step 
scenario would imply a sizable modification
of the power-law dependence (intermittency) 
of the normalized factorial moments
versus the phase space (pseudorapidity) binning. 
Determining the
multifractality nature of multiparticle production
could be useful to assess the hypothetical contribution from a HS.
Our main conclusion is that those 
cascades initiated by very massive hidden (un)particles
should become
{\em less multifractal} (= more monofractal or  
merely less fractal)  
than in conventional events, basically because of a larger 
correlation length, leading to
a suggestive (though formal)
resemblance with a second-order phase transition from
QGP to deconfined hadrons in heavy-ion collisions, 
in spite of very different underlying
physical processes. 

We thereby have worked out a detailed $\lq\lq$exercise'' 
on $M$- and $F$-scaling of factorial moments,  
applying both L\'evy law and 
Ginzburg-Landau descriptions, commonly used in the
search for QGP. 
Several reference values for the $\mu$ and $\nu$ exponents
(related to the cascade character of the production process) 
were obtained as illustrative
examples (not actual predictions).
We concluded that both estimators should
be smaller in a 3-step cascade than in a
2-step process, hopefully providing a quantitative
way of assessing the presence of a HS in the cascade 
once experimental data are employed.

Other techniques to obtain relic information carried out by 
final-state particles about their (grand) parent state, not
addressed in this paper,
should be considered however in a future extension of this work:
azimuthal correlations, rapidity gaps, etc.
Besides, note that correlations should be
stronger in multi-dimensional phase-space \cite{Ochs:1990wa}
than in a lower dimensional projection like rapidity space 
\footnote{This feature is 
pictorially described in \cite{DeWolf:1995pc} 
by comparing the (rather continuous) two-dimensional
shadow of a tree, and the self-similar branching of 
this tree in three-dimensional space.}.

\subsubsection*{Proposal}
Generally speaking, the search strategy for new phenomena 
consists in distinguishing the characteristic signal from
the SM background. Direct signals (e.g. based on jets or
missing energy/$p_{ \bot}$) from a new state of 
non-standard matter (i.e. from a hidden sector) can 
be difficult to observe, 
and/or discriminate among different possibilities yielding
similar phenomenology. Our proposal can thus
be seen as a way of optimizing all
the potential information to be obtained 
from LHC, Tevatron and RHIC experiments:
an example of sinergy.

To this aim, looking back at the impressive
work already done in the analysis of high-energy
hadron-hadron collisions, we have put forward a 
complementary inclusive analysis 
of multiparticle dynamics in inelastic p-p collisions
based on (pseudo)rapidity correlations among
the emitted hadrons. In addition, 
correlations between hadrons and photons
should also be of great interest, especially 
in certain HS models predicting a
large rate of soft photon radiation in events.

Selection (off-line) cuts  should be applied to 
a statistically significant sample of MBE's 
based on criteria like: high multiplicity and sphericity,
jet activity (not necessarily large-$p_{\bot}$ jets) and,
probably, model-dependent criteria such as requiring
heavy flavour, multiple leptons and/or photons in the final state.
Those events not passing the cuts should play
the role of the null hypothesis in a hypothesis test 
\cite{Frodesen} on the existence of a HS manifesting in semi-hard
physics.

Needless to say, a much more complete study including a
realistic modeling of hadroproduction
and simulation of detector effects
is required to decide about the usefulness of this proposal.

\subsection*{Acknowledgments}
I gratefully acknowledge Carlos Garcia-Canal, Emilio Higon, Vicent Martinez,
Vasiliki Mitsou, Lluis Oliver, Carlos Pajares, 
Joannis Papavassiliou and Edward Sarkisyan-Grinbaum for many useful
comments. I also thank MICINN and Generalitat Valenciana for financial support
under research grants FPA2005-01678, FPA2008-02878 and GVPRE/2008/003.
Since this paper refers to a lot of background material and references,
apologies for possible oversights totally unintended.\\


\begin{thebibliography}{999}




\bibitem{Mangano:2008ag}
  M.~L.~Mangano,
  arXiv:0802.0026 [hep-ph];
  M.~L.~Mangano,
  arXiv:0809.1567 [hep-ph].



\bibitem{Harnik:2008ax}
  R.~Harnik and T.~Wizansky,
  arXiv:0810.3948 [hep-ph].

\bibitem{Giovannini:2003ft}
  A.~Giovannini and R.~Ugoccioni,
  Phys.\ Rev.\  D {\bf 68} (2003) 034009
  [arXiv:hep-ph/0304128].

\bibitem{lhc}
  G.~Aad {\it et al.}  [ATLAS Collaboration],
  JINST {\bf 3} (2008) S08003;
  R.~Adolphi {\it et al.}  [CMS Collaboration],
  JINST {\bf 3} (2008) S08004;
  A.~A.~Alves {\it et al.}  [LHCb Collaboration],
  JINST {\bf 3} (2008) S08005.

\bibitem{alice}
  K.~Aamodt {\it et al.}  [ALICE Collaboration],
  JINST {\bf 3} (2008) S08002.


\bibitem{Wells:2008xg}
  J.~D.~Wells,
  arXiv:0803.1243 [hep-ph].



\bibitem{Jones:2004dv}
  B.~J.~T.~Jones, V.~J.~Martinez, E.~Saar and V.~Trimble,
  Rev.\ Mod.\ Phys.\  {\bf 76} (2005) 1211
  [arXiv:astro-ph/0406086].



\bibitem{Tannenbaum:2005by}
  M.~J.~Tannenbaum,
  J.\ Phys.\ Conf.\ Ser.\  {\bf 27} (2005) 1
  [arXiv:nucl-ex/0507020].


\bibitem{Hwa:1992uq}
  R.~C.~Hwa and M.~T.~Nazirov,
  Phys.\ Rev.\ Lett.\  {\bf 69} (1992) 741;
  R.~C.~Hwa and J.~c.~Pan,
  Phys.\ Lett.\  B {\bf 297} (1992) 35;
  R.~C.~Hwa,
  Phys.\ Rev.\  D {\bf 47} (1993) 2773.


\bibitem{Dremin:1977wc}
  I.~M.~Dremin and C.~Quigg,
  Science {\bf 199}, 937 (1978)
  [Usp.\ Fiz.\ Nauk {\bf 124}, 535 (1978)].



\bibitem{Fermi:1950jd}
  E.~Fermi,
  Prog.\ Theor.\ Phys.\  {\bf 5} (1950) 570.

\bibitem{Wilson} K.G.Wilson, Proc. Fourteen Scottish Universities
Summer School in Physics (1973), edited by R.L.~Crawford and
R.~Jennings (Academic Press, New York, 1974).




\bibitem{Belenkij:1956cd}
  S.~Z.~Belenkij and L.~D.~Landau,
  Nuovo Cim.\ Suppl.\  {\bf 3S10} (1956) 15
  [Usp.\ Fiz.\ Nauk {\bf 56} (1955) 309];
  L.~D.~Landau,
  Izv.\ Akad.\ Nauk Ser.\ Fiz.\  {\bf 17} (1953) 51.


\bibitem{Hagedorn:1967ua}
  R.~Hagedorn and J.~Ranft,
  Nuovo Cim.\ Suppl.\  {\bf 6} (1968) 169.

\bibitem{Veneciano} 
S.~Fubini and G.~Veneziano, Nuovo. Cim. A {\bf 64} (1969) 811. 


\bibitem{cabibbo} N. Cabibbo and G. Parisi, Phys.\ Lett.\ B{\bf 59} (1975) 67.



\bibitem{Frazer:1972xs}
  W.~R.~Frazer {\it et al.},
  Rev.\ Mod.\ Phys.\  {\bf 44} (1972) 284.



\bibitem{Meunier:1974nj}
  J.~L.~Meunier and G.~Plaut,
  Nucl.\ Phys.\  B {\bf 87} (1975) 74;
  E.~L.~Berger and G.~C.~Fox,
  Phys.\ Lett.\  B {\bf 47} (1973) 162.




\bibitem{Alver:2007wy}
  B.~Alver {\it et al.}  [PHOBOS Collaboration],
  Phys.\ Rev.\  C {\bf 75} (2007) 054913
  [arXiv:0704.0966 [nucl-ex]].


\bibitem{Ansorge:1988fg}
  R.~E.~Ansorge {\it et al.}  [UA5 Collaboration],
  Z.\ Phys.\  C {\bf 37} (1988) 191.





\bibitem{Andersson:1998ec}
  B.~Andersson, G.~Gustafson, M.~Ringner and P.~J.~Sutton,
  Eur.\ Phys.\ J.\  C {\bf 7} (1999) 251
  [arXiv:hep-ph/9808436].



\bibitem{Liu:2008zzm}
  F.~H.~Liu, Y.~Yuan and M.~Y.~Duan,
  Nucl.\ Phys.\  A {\bf 801} (2008) 154.


\bibitem{Liu:2008zz}
  F.~H.~Liu,
  Nucl.\ Phys.\  A {\bf 810} (2008) 159.






\bibitem{Benecke:1969sh}
  J.~Benecke, T.~T.~Chou, C.~N.~Yang and E.~Yen,
  Phys.\ Rev.\  {\bf 188}, 2159 (1969).



\bibitem{Otterlund:1978bw}
  I.~Otterlund {\it et al.},
  Nucl.\ Phys.\  B {\bf 142} (1978) 445.


\bibitem{Busza:2007ke}
  W.~Busza,
  J.\ Phys.\ G {\bf 35} (2008) 044040
  [arXiv:0710.2293 [nucl-ex]].



\bibitem{Sarkisyan:2005rt}
  E.~K.~G.~Sarkisyan and A.~S.~Sakharov,
  AIP Conf.\ Proc.\  {\bf 828} (2006) 35
  [arXiv:hep-ph/0510191]; 
  E.~K.~G.~Sarkisyan and A.~S.~Sakharov,
  arXiv:hep-ph/0410324.



\bibitem{Akindinov:2007rr}
  A.~Akindinov {\it et al.},
  Eur.\ Phys.\ J.\  C {\bf 50} (2007) 341
  [arXiv:0709.1664 [hep-ph]].




\bibitem{Moraes:2007rq}
  A.~Moraes, C.~Buttar and I.~Dawson,
  Eur.\ Phys.\ J.\  C {\bf 50}, 435 (2007).


\bibitem{Mitrovski:2008hb}
  M.~Mitrovski, T.~Schuster, G.~Graef, H.~Petersen and M.~Bleicher,
  arXiv:0812.2041 [hep-ph].


\bibitem{Bialas:2004kt}
  A.~Bialas and M.~Jezabek,
  Phys.\ Lett.\  B {\bf 590} (2004) 233
  [arXiv:hep-ph/0403254]. 




\bibitem{Capella:1992yb}
  A.~Capella, U.~Sukhatme, C.~I.~Tan and J.~Tran Thanh Van,
  Phys.\ Rept.\  {\bf 236} (1994) 225.



\bibitem{Kittel:2004xr}
  W.~Kittel,
  Acta Phys.\ Polon.\  B {\bf 35} (2004) 2817.


\bibitem{Kittel:2005fu}
  W.~Kittel and E.~A.~De Wolf,
  ``Soft Multihadron Dynamics,''
{\it  World Scientific 2005}


\bibitem{Koba:1972ng}
  Z.~Koba, H.~B.~Nielsen and P.~Olesen,
  Nucl.\ Phys.\  B {\bf 40} (1972) 317.


\bibitem{Kanki:1989cx}
  T.~Kanki, K.~Kinoshita, H.~Sumiyoshi and F.~Takagi,
  Prog.\ Theor.\ Phys.\ Suppl.\  {\bf 97B} (1989) 1.


\bibitem{Bouzas:1992gg}
  A.~O.~Bouzas, L.~N.~Epele, H.~Fanchiotti and C.~A.~Garcia Canal,
  Z.\ Phys.\  C {\bf 56} (1992) 107.



\bibitem{Hegyi:1999aa}
  S.~Hegyi,
  Phys.\ Lett.\  B {\bf 467} (1999) 126;
  S.~Hegyi,
  Phys.\ Lett.\  B {\bf 466} (1999) 380.




\bibitem{Polyakov:1970xd}
  A.~M.~Polyakov,
  JETP Lett.\  {\bf 12} (1970) 381
  [Pisma Zh.\ Eksp.\ Teor.\ Fiz.\  {\bf 12} (1970) 538].



\bibitem{Dremin:2000ep}
  I.~M.~Dremin and J.~W.~Gary,
  Phys.\ Rept.\  {\bf 349} (2001) 301
  [arXiv:hep-ph/0004215].



\bibitem{Dremin:2004ts}
  I.~M.~Dremin and V.~A.~Nechitailo,
  Phys.\ Rev.\  D {\bf 70}, 034005 (2004)
  [arXiv:hep-ph/0402286].


\bibitem{Sarkisian:2000ux}
  E.~K.~G.~Sarkisyan,
  Phys.\ Lett.\  B {\bf 477}, 1 (2000)
  [arXiv:hep-ph/0001262].


\bibitem{Giovannini:1985mz}
  A.~Giovannini and L.~Van Hove,
  Z.\ Phys.\  C {\bf 30} (1986) 391;
  A.~Giovannini and L.~Van Hove,
  Acta Phys.\ Polon.\  B {\bf 19} (1988) 495.


\bibitem{Giovannini:2004yk}
  A.~Giovannini and R.~Ugoccioni,
  Int.\ J.\ Mod.\ Phys.\  A {\bf 20} (2005) 3897
  [arXiv:hep-ph/0405251].



\bibitem{Brambilla:2006zt}
  M.~Brambilla, A.~Giovannini and R.~Ugoccioni,
  arXiv:hep-ph/0605269.



\bibitem{Giovannini:2001kj}
  A.~Giovannini, S.~Lupia and R.~Ugoccioni,
  Phys.\ Rev.\  D {\bf 65} (2002) 094028
  [arXiv:hep-ph/0203205].


\bibitem{Kaidalov:1983ew}
  A.~B.~Kaidalov and K.~A.~Ter-Martirosyan,
  Sov.\ J.\ Nucl.\ Phys.\  {\bf 40} (1984) 135
  [Yad.\ Fiz.\  {\bf 40} (1984) 211].

\bibitem{Andersson:1983ia}
  B.~Andersson, G.~Gustafson, G.~Ingelman and T.~Sjostrand,
  Phys.\ Rept.\  {\bf 97} (1983) 31.


\bibitem{Back:2003xk}
  B.~B.~Back {\it et al.}  [PHOBOS Collaboration],
  arXiv:nucl-ex/0301017;
  B.~B.~Back {\it et al.}  [PHOBOS Collaboration],
  Phys.\ Rev.\  C {\bf 74} (2006) 021902;
  S.~Manly {\it et al.}  [PHOBOS Collaboration],
  Nucl.\ Phys.\  A {\bf 774} (2006) 523
  [arXiv:nucl-ex/0510031].



\bibitem{Cunqueiro:2008uu}
  L.~Cunqueiro, J.~Dias de Deus and C.~Pajares,
  arXiv:0806.0523 [hep-ph].



\bibitem{Dremin:2004zy}
  I.~M.~Dremin,
  Phys.\ Atom.\ Nucl.\  {\bf 68} (2005) 758
  [Yad.\ Fiz.\  {\bf 68} (2005) 790]
  [arXiv:hep-ph/0404202].



\bibitem{Abbiendi:2006qr}
  G.~Abbiendi {\it et al.}  [OPAL Collaboration],
  Phys.\ Lett.\  B {\bf 638} (2006) 30
  [arXiv:hep-ex/0604003].


\bibitem{Sjostrand:1987su}
  T.~Sjostrand and M.~van Zijl,
  Phys.\ Rev.\  D {\bf 36} (1987) 2019.



\bibitem{Sjostrand:2008vc}
  T.~Sjostrand,
  arXiv:0809.0303 [hep-ph].



\bibitem{Lund}
B.~Andersson, {\em The Lund Model} (Cambridge University Press, 1998).



\bibitem{Sjostrand:2007gs}
  T.~Sjostrand, S.~Mrenna and P.~Skands,
  Comput.\ Phys.\ Commun.\  {\bf 178}, 852 (2008)
  [arXiv:0710.3820 [hep-ph]].


\bibitem{Muller:2000zf}
  H.~Muller,
  Eur.\ Phys.\ J.\  C {\bf 18} (2001) 563
  [arXiv:hep-ph/0011350].




\bibitem{Arakelyan:2007kq}
  G.~H.~Arakelyan, C.~Merino, C.~Pajares and Yu.~M.~Shabelski,
  Eur.\ Phys.\ J.\  C {\bf 54} (2008) 577
  [arXiv:0709.3174 [hep-ph]].



\bibitem{Adcox:2004mh}
  K.~Adcox {\it et al.}  [PHENIX Collaboration],
  Nucl.\ Phys.\  A {\bf 757} (2005) 184
  [arXiv:nucl-ex/0410003].



\bibitem{Matsui:1986dk}
  T.~Matsui and H.~Satz,
  Phys.\ Lett.\  B {\bf 178} (1986) 416.


\bibitem{Brambilla:2004wf}
  N.~Brambilla {\it et al.}  [Quarkonium Working Group],
  arXiv:hep-ph/0412158.


\bibitem{Arsene:2004fa}
  I.~Arsene {\it et al.}  [BRAHMS Collaboration],
  Nucl.\ Phys.\  A {\bf 757} (2005) 1
  [arXiv:nucl-ex/0410020].



\bibitem{Amelin:1994mf}
  N.~S.~Amelin, N.~Armesto, M.~A.~Braun, E.~G.~Ferreiro and C.~Pajares,
  Phys.\ Rev.\ Lett.\  {\bf 73} (1994) 2813.


\bibitem{Antoniou:1998ud}
  N.~G.~Antoniou, F.~K.~Diakonos, C.~N.~Ktorides and M.~Lahanas,
  Phys.\ Lett.\  B {\bf 432} (1998) 8.

\bibitem{Kovchegov:1999ep}
  Y.~V.~Kovchegov, E.~Levin and L.~D.~McLerran,
  Phys.\ Rev.\  C {\bf 63} (2001) 024903
  [arXiv:hep-ph/9912367].


\bibitem{Voloshin:2005qj}
  S.~A.~Voloshin,
  J.\ Phys.\ Conf.\ Ser.\  {\bf 50} (2006) 111
  [arXiv:nucl-ex/0505003].


\bibitem{Adams:2006yt}
  J.~Adams {\it et al.}  [STAR Collaboration],
  Phys.\ Rev.\ Lett.\  {\bf 97} (2006) 162301
  [arXiv:nucl-ex/0604018].


\bibitem{Noferini:2007zz}
  F.~Noferini,
  Eur.\ Phys.\ J.\  C {\bf 52} (2007) 247.


\bibitem{McLerran:1993ka}
  L.~D.~McLerran and R.~Venugopalan,
  Phys.\ Rev.\  D {\bf 49} (1994) 3352
  [arXiv:hep-ph/9311205];
  L.~D.~McLerran and R.~Venugopalan,
  Phys.\ Rev.\  D {\bf 49} (1994) 2233
  [arXiv:hep-ph/9309289].



\bibitem{Armesto:2006bv}
  N.~Armesto, L.~McLerran and C.~Pajares,
  Nucl.\ Phys.\  A {\bf 781} (2007) 201.

\bibitem{Kovchegov:2008qn}
  Y.~V.~Kovchegov,
  arXiv:0811.2438 [hep-ph].



\bibitem{McLerran:2008es}
  L.~McLerran,
  arXiv:0812.4989 [hep-ph].


\bibitem{Srivastava:2007aw}
  B.~K.~Srivastava, R.~P.~Scharenberg and T.~J.~Tarnowsky,
  arXiv:nucl-ex/0702040;
  B.~K.~Srivastava  [STAR Collaboration],
  Int.\ J.\ Mod.\ Phys.\  E {\bf 16} (2008) 3371
  [arXiv:nucl-ex/0702054].


\bibitem{Dumitru:2008wn}
  A.~Dumitru, F.~Gelis, L.~McLerran and R.~Venugopalan,
  Nucl.\ Phys.\  A {\bf 810} (2008) 91
  [arXiv:0804.3858 [hep-ph]].



\bibitem{Field:2007zz}
  R.~Field  [CDF Collaboration],
  AIP Conf.\ Proc.\  {\bf 928} (2007) 91.



\bibitem{Skands:2007zz}
  P.~Z.~Skands,
  FERMILAB-CONF-07-706-T.





\bibitem{Georgi:2007ek}
  H.~Georgi,
  Phys.\ Rev.\ Lett.\  {\bf 98} (2007) 221601
  [arXiv:hep-ph/0703260];
  H.~Georgi,
  Phys.\ Lett.\  B {\bf 650} (2007) 275
  [arXiv:0704.2457 [hep-ph]].


\bibitem{Nikolic:2008ax}
  H.~Nikolic,
  Mod.\ Phys.\ Lett.\  A {\bf 23} (2008) 2645
  [arXiv:0801.4471 [hep-ph]].


 
\bibitem{Kikuchi}
 T.~Kikuchi and N.~Okada,
  Phys.\ Lett.\  B {\bf 661}, 360 (2008)
  [arXiv:0707.0893 [hep-ph]].


\bibitem{Sannino}
  F.~Sannino and R.~Zwicky,
  arXiv:0810.2686 [hep-ph].


\bibitem{Fox:2007sy}
  P.~J.~Fox, A.~Rajaraman and Y.~Shirman,
  Phys.\ Rev.\  D {\bf 76} (2007) 075004
  [arXiv:0705.3092 [hep-ph]];
  A.~Delgado, J.~R.~Espinosa and M.~Quiros,
  JHEP {\bf 0710} (2007) 094
  [arXiv:0707.4309 [hep-ph]];
  



\bibitem{Kikuchi2}
 T.~Kikuchi and N.~Okada,
  Phys.\ Lett.\ B {\bf 665}, 186 (2008)
  [arXiv:0711.1506 [hep-ph]].



\bibitem{Davoudiasl:2007jr}
  H.~Davoudiasl,
  Phys.\ Rev.\ Lett.\  {\bf 99} (2007) 141301
  [arXiv:0705.3636 [hep-ph]];
  S.~L.~Chen, X.~G.~He, X.~P.~Hu and Y.~Liao,
  arXiv:0710.5129 [hep-ph];
  B.~Grzadkowski and J.~Wudka,
  arXiv:0809.0977 [hep-ph].


\bibitem{He:2008xv}
  X.~G.~He and L.~Tsai,
  JHEP {\bf 0806} (2008) 074
  [arXiv:0805.3020 [hep-ph]];
  M.~Luo and G.~Zhu,
  Phys.\ Lett.\  B {\bf 659} (2008) 341
  [arXiv:0704.3532 [hep-ph]].


\bibitem{Neubert:2007kh}
  M.~Neubert,
  Phys.\ Lett.\  B {\bf 660} (2008) 592
  [arXiv:0708.0036 [hep-ph]].

\bibitem{Feng:2008ae}
  J.~L.~Feng, A.~Rajaraman and H.~Tu,
  Phys.\ Rev.\  D {\bf 77} (2008) 075007
  [arXiv:0801.1534 [hep-ph]]




\bibitem{Cheung:2008xu}
  K.~Cheung, W.~Y.~Keung and T.~C.~Yuan,
  arXiv:0809.0995 [hep-ph].



\bibitem{Rajaraman:2008bc}
  A.~Rajaraman,
  arXiv:0806.1533 [hep-ph].

\bibitem{Delgado:2008gj}
  A.~Delgado, J.~R.~Espinosa, J.~M.~No and M.~Quiros,
  arXiv:0812.1170 [hep-ph].


\bibitem{Strassler:2006im}
  M.~J.~Strassler and K.~M.~Zurek,
  Phys.\ Lett.\  B {\bf 651}, 374 (2007)
  [arXiv:hep-ph/0604261];
  M.~J.~Strassler,
  arXiv:0801.0629 [hep-ph];
  M.~J.~Strassler,
  arXiv:0806.2385 [hep-ph].

\bibitem{Han:2007ae}
  T.~Han, Z.~Si, K.~M.~Zurek and M.~J.~Strassler,
  JHEP {\bf 0807} (2008) 008
  [arXiv:0712.2041 [hep-ph]];
  M.~J.~Strassler,
  arXiv:hep-ph/0607160.





\bibitem{Patt:2006fw}
  B.~Patt and F.~Wilczek,
  arXiv:hep-ph/0605188.


\bibitem{Chang:2008cw}
  S.~Chang, R.~Dermisek, J.~F.~Gunion and N.~Weiner,
  arXiv:0801.4554 [hep-ph].


\bibitem{Domingo:2008rr}
  F.~Domingo, U.~Ellwanger, E.~Fullana, C.~Hugonie and M.~A.~Sanchis-Lozano,
  arXiv:0810.4736 [hep-ph].


\bibitem{Belyaev:2008yj}
  A.~Belyaev, R.~Foadi, M.~T.~Frandsen, M.~Jarvinen, A.~Pukhov and F.~Sannino,
  arXiv:0809.0793 [hep-ph].



\bibitem{Kang:2008ea}
  J.~Kang and M.~A.~Luty,
  arXiv:0805.4642 [hep-ph].

\bibitem{Giddings:2001bu}
  S.~B.~Giddings and S.~D.~Thomas,
  Phys.\ Rev.\  D {\bf 65} (2002) 056010
  [arXiv:hep-ph/0106219]; 
  S.~Dimopoulos and G.~L.~Landsberg,
  Phys.\ Rev.\ Lett.\  {\bf 87} (2001) 161602
  [arXiv:hep-ph/0106295].


\bibitem{Foa:1975eu}
  L.~Foa,
  Phys.\ Rept.\  {\bf 22}, 1 (1975);
  S.~R.~Amendolia {\it et al.},
  Phys.\ Lett.\  B {\bf 48}, 359 (1974); 
  R.~Singer {\it et al.},
  Phys.\ Lett.\  B {\bf 49} (1974) 481



\bibitem{Porter:2005gp}
  R.~J.~Porter and T.~A.~Trainor  [STAR Collaboration],
  J.\ Phys.\ Conf.\ Ser.\  {\bf 27} (2005) 98
  [arXiv:hep-ph/0506172].


\bibitem{Simic:1979pp}
  L.~Simic {\it et al.},
  Z.\ Phys.\  C {\bf 2} (1979) 291;
  J.~M.~Bolta, E.~Higon and M.~A.~Sanchis,
  Nuovo Cim.\  A {\bf 59} (1980) 173;
  D.~Ghosh, J.~Roy, M.~Basu, K.~Sengupta, S.~Naha, A.~Bhattacharyya and T.~G.~Thakurta,
  Phys.\ Rev.\  D {\bf 26} (1982) 2983.


\bibitem{Breakstone:1982hg}
  A.~Breakstone {\it et al.}  [ABCDHW Collaboration],
  Phys.\ Lett.\  B {\bf 114} (1982) 383.




\bibitem{Capella:1978rg}
  A.~Capella and A.~Krzywicki,
  Phys.\ Rev.\  D {\bf 18} (1978) 4120.



\bibitem{Abbott:1995as}
  T.~Abbott {\it et al.}  [E-802 Collaboration],
  Phys.\ Rev.\  C {\bf 52} (1995) 2663;
  S.~S.~Adler {\it et al.}  [PHENIX Collaboration],
  Phys.\ Rev.\  C {\bf 76} (2007) 034903
  [arXiv:0704.2894 [nucl-ex]].


\bibitem{Csorgo:2004sr}
  T.~Csorgo, S.~Hegyi, T.~Novak and W.~A.~Zajc,
  Acta Phys.\ Polon.\  B {\bf 36} (2005) 329
  [arXiv:hep-ph/0412243].



\bibitem{Andreev:1995rc}
  I.~V.~Andreev, M.~Biyajima, I.~M.~Dremin and N.~Suzuki,
  Int.\ J.\ Mod.\ Phys.\  A {\bf 10} (1995) 3951
  [arXiv:hep-ph/9501345].




\bibitem{DeWolf:1995pc}
  E.~A.~De Wolf, I.~M.~Dremin and W.~Kittel,
  Phys.\ Rept.\  {\bf 270} (1996) 1
  [arXiv:hep-ph/9508325].



\bibitem{Capella:1989xg}
  A.~Capella, K.~Fialkowski and A.~Krzywicki, LPTHE-ORSAY-89-21,
Conference on Multiparticle Dynamics, La Thuile, Italy, Mar 20-22, 1989.


\bibitem{Abbiendi:2001bu}
  G.~Abbiendi {\it et al.}  [OPAL Collaboration],
  Phys.\ Lett.\  B {\bf 523} (2001) 35
  [arXiv:hep-ex/0110051];
  G.~Abbiendi {\it et al.}  [OPAL Collaboration],
  Eur.\ Phys.\ J.\  C {\bf 11} (1999) 239
  [arXiv:hep-ex/9902021].

\bibitem{Tawfik:2000cn}
  A.~M.~Tawfik,
  arXiv:hep-ph/0012022.



\bibitem{Li:2007zzt}
  J.~S.~Li, F.~H.~Liu and D.~H.~Zhang,
  Chin.\ Phys.\ Lett.\  {\bf 24} (2007) 2789;
  D.~H.~Zhang {\it et al.},
  Chin.\ Phys.\  {i\bf 16} (2007) 2689;
D.~H.~Zhang {\it et al.},
  Chin.\ Phys.\  {\bf 16} (2007) 2683
  [Radiat.\ Meas.\  {\bf 43} (2008) S258].



\bibitem{Dremin:1993fr}
  I.~M.~Dremin,
  Mod.\ Phys.\ Lett.\  A {\bf 8} (1993) 2747.



\bibitem{Chau:1992uq}
  L.~L.~Chau and D.~W.~Huang,
  Phys.\ Lett.\  B {\bf 283} (1992) 1.




\bibitem{Bialas:1985jb}
  A.~Bialas and R.~B.~Peschanski,
  Nucl.\ Phys.\  B {\bf 273} (1986) 703;
  A.~Bialas and R.~B.~Peschanski,
  Nucl.\ Phys.\  B {\bf 308} (1988) 857.


\bibitem{Carruthers:1989iw}
  P.~Carruthers and I.~Sarcevic,
  Phys.\ Rev.\ Lett.\  {\bf 63}, 1562 (1989)
  [Erratum-ibid.\  {\bf 63}, 2612 (1989)].




\bibitem{Ochs:1988ky}
  W.~Ochs and J.~Wosiek,
  Phys.\ Lett.\  B {\bf 214} (1988) 617.




\bibitem{Carruthers:1983zx}
  P.~Carruthers and M.~Duong-Van, LA-UR-83-2419.




\bibitem{Uhlig:1977dc}
  S.~Uhlig, I.~Derado, R.~Meinke and H.~Preissner,
  Nucl.\ Phys.\  B {\bf 132} (1978) 15.



\bibitem{Alexander:2000ux}
  G.~Alexander and E.~Sarkisyan,
  Phys.\ Lett.\  B {\bf 487} (2000) 215
  [arXiv:hep-ph/0005212];
  G.~Alexander and E.~Sarkisyan,
  Nucl.\ Phys.\ Proc.\ Suppl.\  {\bf 92} (2001) 211
  [arXiv:hep-ph/0008174].


\bibitem{Berger:1974vn}
  E.~L.~Berger,
  Nucl.\ Phys.\  B {\bf 85} (1975) 61.


\bibitem{Bell:1983di}
  W.~Bell {\it et al.}  [CERN-Heidelberg-Lund Collaboration],
  Z.\ Phys.\  C {\bf 22} (1984) 109.


\bibitem{Cunqueiro:2006xe}
  L.~Cunqueiro, E.~G.~Ferreiro and C.~Pajares,
  PoS C {\bf FRNC2006} (2006) 019
  [arXiv:hep-ph/0611034];
  J.~Dias de Deus, C.~Pajares and C.~A.~Salgado,
  Phys.\ Lett.\  B {\bf 407} (1997) 335
  [arXiv:hep-ph/9702398].



\bibitem{Abreu:2007kv}
  N.~Armesto {\it et al.},
  J.\ Phys.\ G {\bf 35} (2008) 054001
  [arXiv:0711.0974 [hep-ph]]




\bibitem{Brogueira:2006yk}
  P.~Brogueira and J.~Dias de Deus,
  Phys.\ Lett.\  B {\bf 653} (2007) 202
  [arXiv:hep-ph/0611329].




\bibitem{Brogueira:2007ub}
  P.~Brogueira, J.~Dias de Deus and J.~G.~Milhano,
  Phys.\ Rev.\  C {\bf 76} (2007) 064901



\bibitem{Giovannini:2002za}
  A.~Giovannini and R.~Ugoccioni,
  Phys.\ Rev.\  D {\bf 66} (2002) 034001
  [arXiv:hep-ph/0205156].



\bibitem{Dremin:1989wh}
  I.~M.~Dremin,
FERMILAB-PUB-89-071-T.


\bibitem{Albajar:1992hr}
  C.~Albajar {\it et al.}  [UA1 Collaboration],
  Z.\ Phys.\  C {\bf 56} (1992) 37.



\bibitem{Sarkisian:1994xa}
  E.~K.~Sarkisian, L.~K.~Gelovani, G.~L.~Gogiberidze and G.~G.~Taran,
  Phys.\ Lett.\  B {\bf 347} (1995) 439
  [arXiv:hep-ph/9412336].


\bibitem{Ahmad:2006sb}
  S.~Ahmad and M.~A.~Ahmad,
  J.\ Phys.\ G {\bf 32} (2006) 1279.


\bibitem{Dahlqvist:1989yc}
  P.~Dahlqvist, B.~Andersson and G.~Gustafson,
  Nucl.\ Phys.\  B {\bf 328} (1989) 76;
  B.~Andersson, G.~Gustafson, A.~Nilsson and C.~Sjogren,
  Z.\ Phys.\  C {\bf 49} (1991) 79;
  G.~Gustafson and A.~Nilsson,
  Nucl.\ Phys.\  B {\bf 355} (1991) 106;
  G.~Gustafson and A.~Nilsson,
  Z.\ Phys.\  C {\bf 52} (1991) 533.



\bibitem{Ochs:1992tg}
  W.~Ochs and J.~Wosiek,
  Phys.\ Lett.\  B {\bf 305} (1993) 144.



\bibitem{Lipa:1989yh}
  P.~Lipa and B.~Buschbeck,
  Phys.\ Lett.\  B {\bf 223} (1989) 465.


\bibitem{Hwa:1989vn}
  R.~C.~Hwa,
  Phys.\ Rev.\  D {\bf 41} (1990) 1456.


\bibitem{Bialas:1990xd}
  A.~Bialas and R.~C.~Hwa,
  Phys.\ Lett.\  B {\bf 253} (1991) 436.




\bibitem{Brax:1990jv}
  P.~Brax and R.~B.~Peschanski,
  Phys.\ Lett.\  B {\bf 253} (1991) 225.





\bibitem{Zhang:1995uf}
  Y.~Zhang, L.~S.~Liu and Y.~f.~Wu,
  Z.\ Phys.\  C {\bf 71} (1996) 499.

\bibitem{Tawfik:2001az}
  A.~M.~Tawfik,
  arXiv:hep-ph/0104004;
  A.~M.~Tawfik and E.~Ganssauge,
  Heavy Ion Phys.\  {\bf 12} (2000) 53
  [arXiv:hep-ph/0012008].

\bibitem{Bialas:1992ca}
  A.~Bialas,
  Acta Phys.\ Polon.\  B {\bf 23} (1992) 561.


\bibitem{Utyuzh:1999eb}
  O.~V.~Utyuzh, G.~Wilk and Z.~Wlodarczyk,
  Czech.\ J.\ Phys.\  {\bf 50S2} (2000) 132
  [arXiv:hep-ph/9910355].


\bibitem{Ginzburg:1950sr}
  V.~L.~Ginzburg and L.~D.~Landau,
  Zh.\ Eksp.\ Teor.\ Fiz.\  {\bf 20} (1950) 1064.



\bibitem{Cao:1996tg}
  Z.~Cao, Y.~Gao and R.~C.~Hwa,
  Z.\ Phys.\  C {\bf 72} (1996) 661
  [arXiv:nucl-th/9601011].





\bibitem{Ochs:1990wa}
  W.~Ochs,
  Z.\ Phys.\  C {\bf 50} (1991) 339;
  A.~De Angelis, P.~Lipa and W.~Ochs,
Conf.Proc.C910725V1:724-737,1991.
  


\bibitem{Frodesen} 
A.G.~Frodesen et al., {\em Probability and Statistics
in Particle Physics} (Universitetsforlaget, 1979)



\end{thebibliography}
\end{document}